\documentclass[fleqn]{aa}  
\usepackage{supertabular,booktabs}  
\usepackage{graphicx}
\usepackage{xcolor}
\usepackage{hyperref}
\usepackage{txfonts}
\usepackage{calligra}
\usepackage{calrsfs}
\usepackage[T1]{fontenc}
\usepackage[mathcal]{eucal}
\usepackage{longtable}
\usepackage{multicol}

\usepackage[switch, modulo]{lineno}

\sfcode`A=1000 \sfcode`B=1000 \sfcode`C=1000 \sfcode`D=1000
\sfcode`E=1000 \sfcode`F=1000 \sfcode`G=1000 \sfcode`H=1000
\sfcode`I=1000 \sfcode`J=1000 \sfcode`K=1000 \sfcode`L=1000
\sfcode`M=1000 \sfcode`N=1000 \sfcode`O=1000 \sfcode`P=1000
\sfcode`Q=1000 \sfcode`R=1000 \sfcode`S=1000 \sfcode`T=1000
\sfcode`U=1000 \sfcode`V=1000 \sfcode`W=1000 \sfcode`X=1000
\sfcode`Y=1000 \sfcode`Z=1000

\newcommand{\Msun}{\hbox{M$_\sun$}}
\newcommand{\Rsun}{\hbox{R$_\sun$}}

\newcommand{\zs}{$z_{\rm spec}$}
\newcommand{\zg}{$z_{\rm geo}$}

\newcommand{\wslap}{\texttt{WSLAP+}\,}

\newcommand{\ma}{$^1$}
\newcommand{\mb}{$^2$}
\newcommand{\mc}{$^3$}
\newcommand{\md}{$^4$}
\newcommand{\me}{$^5$}
\newcommand{\mf}{$^6$}

\begin{document}

   \title{Hedorah, the first yellow supergiant Kaiju star candidate at $z=3.7$ revealed by JWST behind AS1063.}
   \titlerunning{Hedorah \& AS1063 lensing}
   \author{J.M. Diego \inst{1}\fnmsep\thanks{jdiego@ifca.unican.es}
       \and J.M. Palencia \inst{1}  
       \and C. Goolsby \inst{2}
       \and C.J. Conselice \inst{2}
       \and D. Perera \inst{3}
       \and L.L.R. Williams \inst{3} 
       \and D.J. Lagattuta \inst{4}
       \and G. Mahler\inst{5}      
       \and J. Richard\inst{6}
       \and K. Sharon \inst{7}      
       \and M. Struble \inst{8}
    }      
   \institute{Instituto de F\'isica de Cantabria (CSIC-UC). Avda. Los Castros s/n. 39005 Santander, Spain 
        \and Jodrell Bank Centre for Astrophysics, Alan Turing Building, University of Manchester, Oxford Road, Manchester M13 9PL, UK 
        \and School of Physics and Astronomy, University of Minnesota,
116 Church Street SE, Minneapolis, MN 55455, USA 
        \and Centre for Astrophysics Research, Dep. of Physics, Astronomy and Mathematics, Univ. of Hertfordshire, Hatfield AL10 9AB, UK 
        \and STAR Institute, Quartier Agora - All\'ee du six Août, 19c B-4000 Liège, Belgium. 
        \and Universit\'e Claude Bernard Lyon 1, CRAL UMR5574, ENS de Lyon, CNRS, Villeurbanne, F-69622, France 
        \and Department of Astronomy, University of Michigan, 1085 S. University Ave, Ann Arbor, MI 48109, USA  
        \and Department of Physics and Astronomy, University of Pennsylvania, 209 South 33rd Street, Philadelphia, PA 19104, USA 
          }
 \abstract{
 We present a new free-form lens model for the $z=0.348$ galaxy cluster AS1063, based on previously spectroscopically confirmed lensed galaxies and new JWST images from the GLIMPSE program. We use the ultra-deep JWST data to identify new counterimages for previously confirmed (spectroscopically) lensed systems. We use the full set of spectroscopically confirmed systems to derive a new lens model, which is later used to confirm many of the previous lensed system candidates and discover new lensed system candidates in the JWST images. We compute the geometric redshifts, time delays, and magnification for all counterimages (confirmed and not confirmed). Among the new systems, and based on photometry, we find a peculiar multiply lensed galaxy with a strong emission line at $\approx 4\, \mu$m that likely corresponds to H$-\beta$ and/or OIII at $z\approx 7.5$. This galaxy could be a little-red-dot or an extreme emission line galaxy. We also identify a yellow supergiant lensed star candidate at $z\approx 3.7$. This star shows some similarities with previous Kaiju stars and we nickname it "Hedorah", in honor of the famous yellow-eyed Kaiju. Previous lensed stars at $z>0.1$ are either blue supergiants or red supergiants, making Hedorah the first yellow supergiant discovered beyond $z=0.1$ and confirming that, despite their rarity, they can also be found at these redshifts. Since many Cepheid stars are yellow supergiants, we consider the possibility that Hedorah could also be the first Cepheid discovered at cosmological distances, but we conclude that Hedorah is more likely a hypergiant yellow star approaching the end of its life.  Alternatively, Hedorah could be a small group of stars, although this is less likely based on Hedorah's peculiar colors and additionally may require the more exotic fuzzy dark matter to help explain the lack of counterimage.
   }

   \maketitle
%
\section{Introduction}


Data from the \textit{James Webb} space telescope (JWST) on galaxy clusters is allowing us to study these massive structures and the background galaxies they magnify with unprecedented detail. It has also enabled a wealth of studies on very young galaxies including detailed spectroscopy for some of them \citep{Adams2023,Bouwens2023,Bunker2023,Adams2024,Atek2024,Carniani2024,PerezGonzalez2025,Conselice2025}, the discovery of many very compact v-shaped spectrum galaxies known as little-red-dots (LRD) \citep{Furtak2023,Barro2024,Matthee2024}, transient events from lensed supernovae (SNe) \citep{Frye2024,Pierel2024,Suyu2025}, or even lensed stars at $z>1$ \citep{Meena2023,Diego2023a,Diego2023c,Furtak2024,Fudamoto25}. On the cluster itself, the high resolution from JWST is ideal to detect unresolved globular clusters which exist in large numbers in massive galaxy clusters, but are much harder to detect with the Hubble Space Telescope (HST) \citep{Lee2022,Faisst2022,Dornan2023,Harris2023,Berkheimer2025}. 

Relevant for this work, one of the areas where JWST data has been revolutionary is in the detection of multiply imaged lensed galaxies. The increased depth and spatial resolution of the new data facilitates the identification of the families of counterimages of lensed galaxies. When combined with the HST data, the rich color information makes identification even more robust. In recent years, some of the new families of images correspond to LRDs, which are expected to harbor a super massive black hole (SMBH) at their center. Their true nature is still debated with recent work suggesting the central engine is a relatively small SMBH surrounded by a dense gas envelope that behaves as the outer layer of a star. Because of this, it is  often referred to as a BH star, or BH*  \citep{Inayoshi2025,deGraaff2025a,Golubchik2025,Rusakov2026}, and is possibly the result of a direct collapse of a massive HI region \cite{Pacucci2026}. These objects are intriguing and may represent the intermediate step between a SMBH seed and the final stage of a SMBH observed in quasars (QSO) or active galactic nucleai (AGN). Multiply lensed LRDs are of particular interest because they can be used to test for their variability (with different scales predicted by different models) and also the magnification allows for higher quality spectra. Very recently it was suggested that LRDs may be pulsating via a mechanism similar to the kappa mechanism in Cepheid stars \citep{Zhang2025}. 
Only better data and more examples of LRDs can help answer this and other questions about these enigmatic objects. 

Perhaps one of the most surprising discoveries in recent years is the ability of  galaxy cluster lenses to magnify individual stars to beyond the detection limit of telescopes such as HST \citep{Kelly2018,Chen2019,Diego2022_Godzilla,Welch2022,Kelly2023}.  
The better resolution of JWST, and its higher sensitivity to the IR allows us to detect more microlensing events from stars in the lensed galaxies that are crossing a caustic. So far, JWST has detected dozens of lensed stars \citep{Meena2023,Diego2023a,Diego2023c,Furtak2024,Fudamoto25} and will discover many more in the near future. Most of these stars were missed by HST because of their very red color (red supergiants) that combined with their redshift makes them undetectable by HST. Detecting these luminous stars beyond $z=1$ is interesting not only because they give us information about how the universe formed the most massive stars when it was much younger \citep{Li2025,Meena2025,Palencia2025b,Williams2026}, but also because these stars can be used as pencil beams scanning  the small scale fluctuations in the lensing potential. \citep{Venumadhav2017,Diego2018,Dai2020,Williams2024,Diego2024b,Ji2025}. In fact, most of the lensed stars discovered so far is thanks to these small variations from microlenses in the intracluster medium, that momentarily increase the magnification of the background star making them appear as transient events (easy to identify in difference images). So far most of the stars discovered by JWST are red supergiants, where HST discovered mostly blue supergiants. No yellow supergiant has been detected yet, which remain elusive due to their rarity. However, these yellow supergiants are of special interest because many of them are Cepheid stars, a standard candle. 

 Finding Cepheids at high redshift would allow for an independent constraint on $H_0$ that does not rely on SNe. However, such a study would not be free of systematic effects. Chief among them is the unknown magnification from the macromodel, but also the stochastic distortion in the magnification introduced by the ubiquitous microlenses (specially from the intracluster medium, but also anywhere else along the line of sight). \cite{Diego2026a} studies in detail the possibility of using Cepheids at $z\approx1$ as standard candles and concludes that the contribution from microlenses can not be ignored. At higher redshifts the prospects of using a lensed Cepheid as a standard candle improve since at these redshifts, the critical curve (CC) moves farther away from the cluster center, and the contribution from microlenses diminishes. Finding a Cepheid at redshift $z>1$ would represent then, not only a great discovery, but a unique opportunity for cosmology. 
 
Among the most promising clusters to study high redshift galaxies and search for elusive microlensing events at high redshift, 
AS1063 \citep[z=0.348, and also known as RXC J2248.7-4431][]{Guzzo2009} is a remarkable example given the depth of JWST data and relatively lower redshift (distance modulus 41.32), when compared with other prominent cluster lenses. This cluster was selected as one of the clusters for the Hubble Frontier Fields (HFF) program \citep{Lotz2017} due to its large mass and existence of known gravitationally lensed galaxies \citep{Postman2012a,Balestra2013,Moona2014,Johnson2014,Richard2014,Zitrin2015}.
As part of this program,  the central $\sim 10$ arcmin$^2$ region was observed by HST in wavelengths ranging from 0.45 $\mu$m to 1.6 $\mu$m and to a depth of $\sim 28.5$ mag in the visible and IR bands. The area observed by the HFF program around AS1063 was later doubled thanks to the Beyond Ultra-deep Frontier Fields and Legacy Observations (BUFFALO) program \citep{Steinhardt2020}, although with shallower observations than in the HFF program. 
JWST data complements the high-quality HST data extending the wavelength coverage to $\approx 5\, \mu$m and reaching greater depths. 
The high-quality data provided by the HFF program allowed the identification of hundreds of images of strongly lensed galaxies behind the six HFF clusters. In AS1063, many of the new lensed galaxies were previously spectroscopically confirmed from the ground \citep{Boone2013,Bradley2014,Moona2014,Johnson2014,Richard2014}, and afterwards with the Multi Unit Spectroscopic Explorer at VLT \citep[MUSE, ][]{Bacon2010} playing a pivotal role \citep{Caminha2016}. The new JWST data, specially the ultra deep data gathered as part of the GLIMPSE program \citep{Atek2025}, reaching a record breaking 5$\sigma$ 30.9 mag depth (point source and aperture diameter of $0".2)$), makes AS1063 an ideal object for lensing studies, especially thanks to its exceptional depth. GLIMPSE data has been used for a variety of studies, from high redshift galaxies \citep{Kokorev2025a,Kokorev2025b,Korber2025,Chemerynska2025}, including very low metallicity ones  \citep[or Pop III galaxy candidates][]{Fujimoto2025}, to early SMBH \citep{Fei2025}. 
The high-resolution JWST data can be used also for the identification of new counterimages and improvement on the lens model for this cluster, or a detailed study in the distribution of globular clusters (GC) tracing the projected distribution of mass (mostly dark matter). GCs are expected to react to the gravitational potential in a very similar fashion as dark matter does. Using GCs as visible tracers of dark matter shall open new avenues to advance on the understanding of this mysterious substance \citep{RamosAlmendares2020,ReinaCampos2022,Diego2023d,Martis2024,Diego2024a,EuclidGC2025a,EuclidGC2025b}.

%
\begin{figure*} 
  \includegraphics[width=\linewidth]{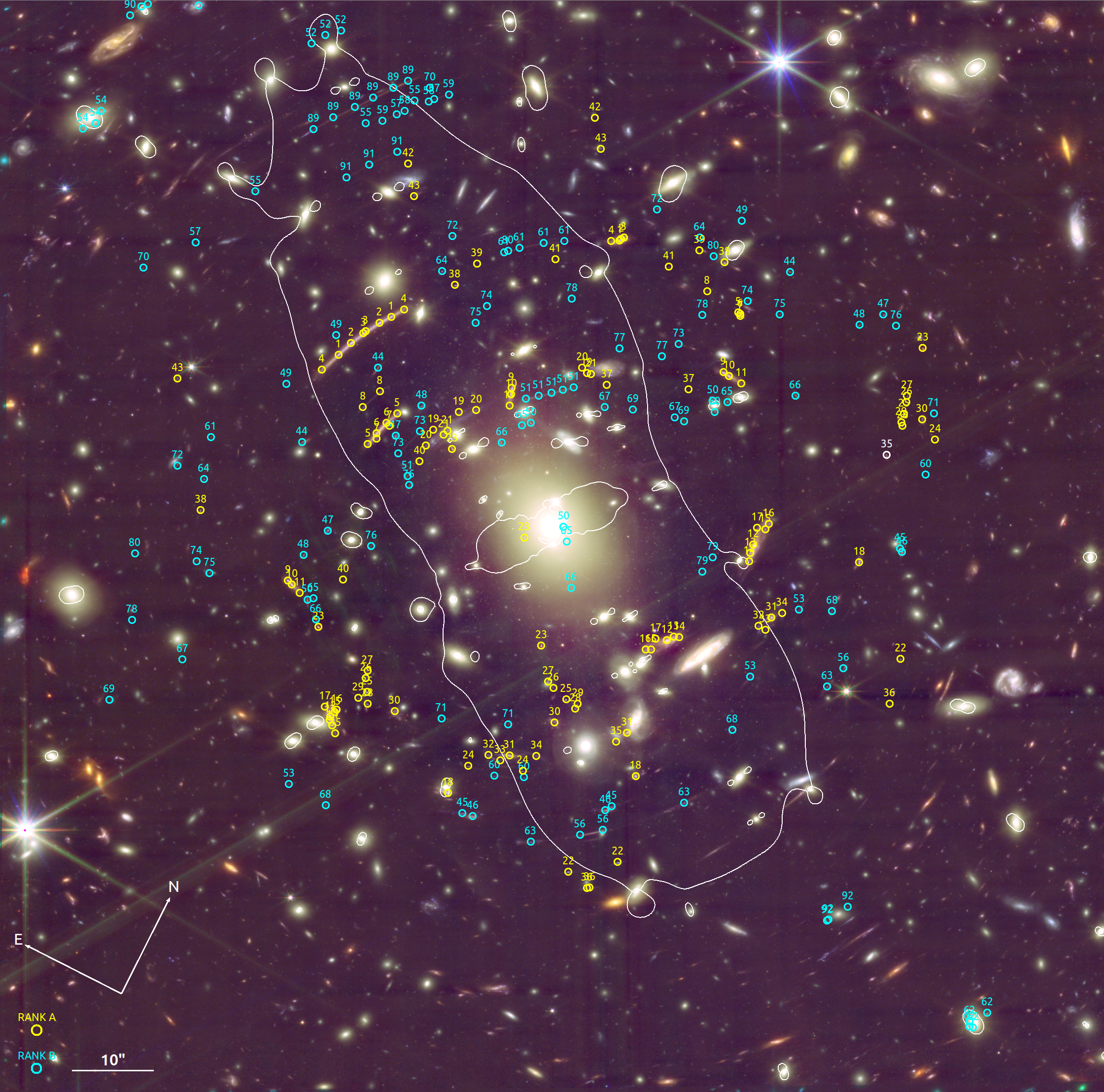} 
      \caption{Central region of AS1063 ($2'.25\times2'.25$) combining 12 HST and JWST filters (range 0.4--5 micron). Yellow circles show the spectroscopically confirmed lensing constraints, or rank A constraints (see Table~\ref{tab_arcs}),  used to derive the lens model. Cyan dots show the rank B constraints (Table~\ref{tab_arcs}) not used in the lens model but highly consistent with it and for which we also derive geometric redshifts, time delays and magnification. The critical curve corresponds to $z_s=3$.
         }
         \label{Fig1}
\end{figure*}

In this paper we revisit this cluster presenting a new hybrid lens model derived using the most up-to-date lensing constraints of AS1063. We use the lens model to unveil new counterimages from previously known systems, confirm lensed system candidates, and identify new lens system candidates. We discuss in more detail several notorious lensed galaxies.  
The first is a LRD or extreme emission line galaxy (EELG) at $z\approx 7.5$. The second one is a highly magnified galaxy that was suspected of crossing a caustic but we reinterpret it as a rare singly lensed but highly magnified galaxy. The remaining two galaxies are caustic crossing galaxies, one at low redshift ($z=0.73$) and one at high redshift ($z\approx 3.7$\footnote{After the original submission of this article, the redshift of this galaxy has been spectroscopically confirmed by Furtak. et al (2026),  $z=3.72$}. The first one at $z=0.73$ shows several candidate red supergiant lensed stars near the cluster CC but that require a second deep observation of this cluster with the same filter to be confirmed. The second caustic crossing galaxy at $z\approx 3.7$ hosts a candidate lensed star (nicknamed Hedorah, and the main focus of this paper) with photometry consistent with being a yellow supergiant star (similar to a Cepheid) or a yellow hypergiant. If confirmed would be the first of its kind at cosmological distances. \\

This paper is organized as follows. 
Sect.~\ref{sec_data} describes the HST and JWST data used in this work. 
This section also includes  a discussion of the lensing constraints derived from the literature and used to derive the lens model. Sect.~\ref{sec_WSLAPplus} describes very briefly the algorithm \wslap used to derive the lens model and presents results from the lens model, in particular to interpret two highly magnified galaxies as well as an LRD/EELG multiply lensed galaxy at $z\approx 7.5$.   
Section~\ref{sec_Hedorah} presents Hedorah, the first yellow supergiant (or hypergiant) star candidate at $z>1$. 
We discuss  our results in  Sect.~\ref{sec_discussion}. Finally, Sect.~\ref{sect_conclusions} summarizes our conclusions. We adopted a standard flat cosmological model with $\Omega_m=0.3$ and $h=0.7$. At the redshift of the lens ($z_l=0.348$), and for this cosmological model, one arcsecond corresponds to 4.921~kpc. For the redshift of the lens and a source at $z_s=3$, the critical surface mass density is $\Sigma_{\rm crit}(z_l,z_s)=2.086\times10^9\, \Msun\,{\rm kpc}^{-2}$. This quantity enters in the definition of the convergence of the lens model, $\kappa(z_l,z_s)=\Sigma/\Sigma_{\rm crit}(z_l,z_s)$.

{\section{HST data, JWST data, and lensing constraints}
\label{sec_data}
AS1063 was observed in September 2024 by JWST as part of the Cycle2 GLIMPSE program (PI Hakim Atek, Program ID G0-3293). An overview of the GLIMPSE program and details of the data are given in \cite{Atek2025}. Typical $5\sigma$ depth for a point source ($0\farcs2$ aperture diameter) is an impressive 30.9 across filters (with exposures varying from filter to filter to reach this depth), resulting in one of the deepest (if not the deepest) images taken with JWST. NIRCam imaging observations were obtained in 9 filters: F090W, F115W, F150W, F200W, F277W, F356W, F410M, F444W, F480M totaling 120 hours of exposure time. 

This JWST data were reduced and processed in a procedure similar to that described in the EPOCHS series \citep[eg][]{conselice2024, Adams_2025}. We use a modified version of the official JWST pipeline version 1.8.2, using Calibration Reference Data System (CRDS) pmap1364 to a pixel scale of 0\farcs03 pix$^{-1}$. To better preserve the light from the foregorund cluster and prevent over-subtraction due to the bright central galaxy we do not include 1/f subtraction, and we employ a custom 2d background subraction between stages 2 and 3. We further include the subtraction of custom wisp templates \citep{Adams_2025}, and the images are aligned to GAIA DR3 during stage 3 \citep{GAIADR3}.

Ancillary archival HST imaging  was also incorporated from the Hubble Legacy Fields \citep{Illingworth2016HLF, Whitaker2019HLF}. We retrieved the latest v2.5 mosaics from MAST for the relevant ACS/WFC filters: F435W, F606W, and F814W on a $0.\!''06$ pixel scale. We reprojected the ACS/WFC mosaics onto the NIRCam World Coordinate System (WCS) and a  $0.\!''03$ pixel scale using the \texttt{reproject} package \citep{2020ascl.soft11023R}.

This work focuses only on the cluster core field, which is the only one characterized by lensing constraints. A portion of the area observed by JWST in this cluster core region is shown in Fig.~\ref{Fig1}. This image (and other images in this paper) is a color composition made after combining previous HST data from the HFF program with the new JWST data. In particular, the blue channel contains HST's F435W, F606W, and F814W, the green channel is a combination of JWST's F090W, F115W, F150W, and F200W, and the red channel includes F277W, F356W, F410M, F444W, and F480M. Between HST's F435W and JWST's F480M there is a factor $\approx 12$ in wavelength (0.4--5 $\mu$m). Computing the effective depth of the stacked images is beyond the scope of this paper, but given the depth of the individual GLIMPSE images it is safe to assume that the stacked image is deeper than AB 31.5.  

The initial set of lensing constraints (rank A) were compiled by the BUFFALO collaboration and come from a compilation of systems identified in earlier work with contributions from different authors (both for the identification of the lensed systems and securing their redshifts). In addition to the BUFFALO data set,  three spectroscopic systems were recently published in \cite{Beauchesne2024}. 
A second set of lensing constraints (rank B, or candidates) were derived after the lens model was constructed from the rank A systems. Many of these are previous candidates that we can confidently confirm with the new JWST data (better color information and spatial resolution than in previous efforts) and validated by the lens model. A few of the previous candidates did not meet one of the criteria (consistency with the lens model and consistency in morphology and/or color among counterimages) and were discarded. In addition to these, the new JWST images reveal new entire systems and new counterimages for some of the previous rank A or rank B systems. We include all rank A and B systems in our curated compilation and make them public together with the lens model products (see data availability at the end of this paper). The final data set of constraints are listed and described in Appendix~\ref{sc_appendixA}. Only constraints with rank A are used to derive the lens model. Constraints of type rank A and type rank B are shown in Fig.~\ref{Fig1} in yellow and cyan, respectively. A few rank B constraints are outside the image shown in  Fig.~\ref{Fig1}, in an area to the NE that is only covered by the BUFFALO data. Spectroscopic redshifts (when available), together with lens model predicted geometric redshifts, time delays and magnifications are listed in Appendix~\ref{sc_appendixA}. 

\section{Lens modeling}\label{sec_WSLAPplus}
Using the rank A constraints described in the previous section, we derived the lens model with the code \wslap \citep{Diego2005,Diego2007,Sendra2014,Diego2016,Diego2026b}. This is a hybrid type of lens modeling technique that combines a free-form decomposition of the smooth, large-scale component (DM, gas, and lower-density stellar contribution to the intracluster light) with a small-scale contribution from cluster galaxies (higher density stellar component in member galaxies). The code combines weak (when available) and strong lensing in a natural way with the large-scale and small-scale components in the mass equally affecting the strong- and weak-lensing observables. For the purposes of this work, only strong-lensing constraints were used.

The algorithm assumes that the smooth mass is described by a superposition of Gaussian functions distributed over a predetermined grid (regular or adaptive), while the compact mass component follows the distribution of light around selected massive member galaxies which are identified from the MUSE spectroscopic sample and red-sequence. The members can be distributed in different groups or layers, each one assuming a constant mass-to-light ratio. For the case of AS1063, we assumed two layers. The first layer contains only the central BCG while the second layer contains the most massive member galaxies ($\sim 100$). The grid for the Gaussians is originally a regular grid of $16\times16$ points where Gaussians of equal width are placed. The uniform grid is equivalent to using no prior on the mass distribution that can appear anywhere in the lens plane. A first solution is derived with this configuration. This first solution is later used to define a new (adaptive) grid which increases the resolution of the grid (and reduces the widths of the Gaussians accordingly) in the region where the first solution contains most mass.  This effectively sets a prior on the derived solution favoring models with more mass wherever the resolution of the grid is higher. The adaptive grid contains 412 points or Gaussians, with a higher concentration at the position of the BCG. Hence the model has 412 free parameters from the grid, 2 free parameters from the layers and $43\times2$  for the $x$ and $y$ positions of each of the 43 source positions corresponding to the rank A lensing constraints in Table~\ref{tab_arcs}. The larger number of free parameters than constraints is not a problem since the optimization is done over a quadratic function that reduces the dimensionality of the problem to a dimension similar to the number of constraints \citep[see][for details]{Diego2007,Diego2026b}. 
The refined solution increases also the number of iterations in the optimization process (from $\sim$10,000 in the first solution with the regular grid to $\sim$200,000 with the adaptive grid).

\begin{figure} 
  \includegraphics[width=\linewidth]{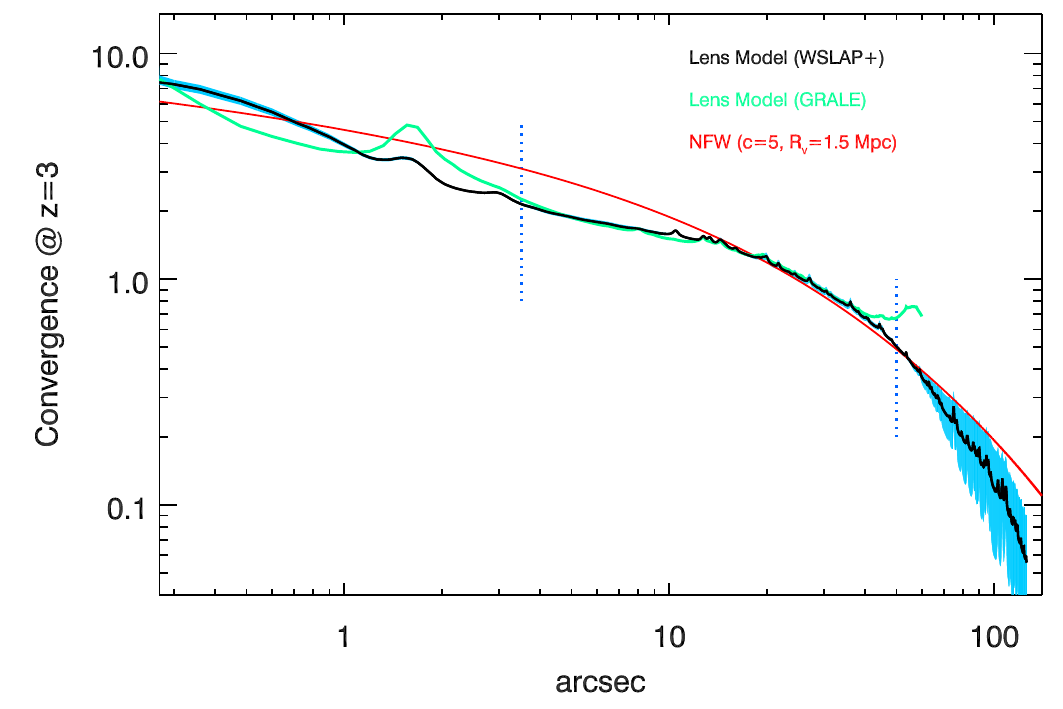} 
      \caption{Mass profile in the central region of AS1063 and centered on the BCG. The projected mass from the lens model is shown in black (in units of $\kappa=\Sigma/\Sigma_{\rm crit}$ at $z_s=3$) with the two blue vertical dotted lines marking the range of distances with lensing constraints. The blue band shows the uncertainty in the lens model obtained after computing the dispersion of 5 solutions where we vary the initial condition of the optimization. For comparison we show as a green line a second independent lens model derived with hybrid GRALE and using the same lensing constraints. The red line shows a NFW profile with parameters shown in the top right corner.
         }
         \label{Fig_Profiles}
\end{figure}

The derived lens model resembles our earlier model produced with \wslap in \cite{Diego2016b} but is more precise due to the increase in the number of constraints. The mass profile is shown in Fig.~\ref{Fig_Profiles} where we compare it with a classic Navarro-Frenk-White (NFW) profile \citep{NFW} in red, and an alternative lens model derived using the same lensing constraints but the alternative model hybrid GRALE \citep{Perera2024,Perera2025}. The blue band represents the uncertainty in our lens model due to the initial guess for the optimization. The final solution suffers of a memory effect in regions which are poorly constrained by the data, remembering the values of the initial guess (a random variable). This can be appreciated in the outer regions, which show the largest dispersion, and where the Gaussian masses are not constrained by the lensing constraints in the inner region. The range of radii constrained by our lensing constraints is shown as two vertical dotted lines. In this range the derived lens model agrees reasonably well with a NFW profile of concentration $c=6$ (typical of massive clusters) and virial radius $R_{\rm v}= 2.7$ Mpc. The limits on the concentration and virial radius are weak since strong lensing alone can not constrain these parameters well. The profile suggests the presence of a large core with radius $r\approx 150$ kpc. The existence of such core is also favored by parametric lens model that allow cored profiles \citep{Bergamini2019,Limousin2022}. 
The departure from the NFW in our profile for radii $r\gtrsim50''$ is typical of \wslap models since the lens model is insensitive to the distribution of mass at large radii which has a much weaker impact on the apparent position of the lensed galaxies in the central region. In these unconstrained regions, the lens model tends to remain close to its initial value before optimization (small values of the mass in the outskirts). Similarly, the profile in the innermost region, $r\lesssim 3''$ is poorly constrained since the lensing constraints are only sensitive to the total mass within this region, and not so much to its radial distribution. 
In our particular case, the profile for $r< 3''$is mostly following the light profile of the compact component in our lens model. The profile shown in  Fig.~\ref{Fig_Profiles} is only one of many possible solutions which are consistent with the lensing constraints. Other valid solutions would look almost identical to this one within the range constrained by lensing (marked with two vertical lines) and would show larger variability in the regions poorly constrained by lensing ($r<3''$ and $r\gtrsim50''$). The focus of this work is not to explore the range of possible solutions, and we are more interested on how one of the valid solutions (for instance the one shown in the figure) helps to interpret some of the galaxies and sources in JWST data. Since the objects of interest fall in the region that is well constrained by lensing, the particular model used for this purpose is not relevant, since all valid models are almost identical in the well constrained region.  

We use the new lens model to confirm previous lensed system candidates and help identify new ones discovered in the JWST data. We also use the lens model to predict the geometric redshift of the system candidates. We follow the method described in \cite{Diego2024a} which basically searches for the redshifts where the lens model focuses each family of counterimages into the most compact configuration possible in the source plane. Similar methods have been followed by other authors. Geometric redshifts are reliable estimations of the true redshift in well calibrated lenses as it is demonstrated in Fig. B.1 of \cite{Diego2024a}. For the particular case of AS1063, our previous model in \cite{Diego2016b} was already successful at correctly predicting the redshift of system 19 in Table~\ref{tab_arcs} (system 10 in our earlier work) before it was measured with MUSE by \cite{Caminha2016}. Due to the bias in the mass in the \wslap mass profile at large radii, we expect some of the geometric redshift predictions from this model to be biased high (to compensate the lower mass) in the rank B systems that fall within these large radii ($r>50"$). These are mostly high-redshift systems ($z>3$) which are the only ones that can appear at such large distances from the cluster center. On the other hand, the rank B systems that appear within the region constrained by rank A systems should have their geometric redshifts well determined, as this is the region of the lens plane that is well calibrated. 

Using the lens model we compute also the magnification and time delays for each of the counterimages using the spectroscopic (when available) or geometric redshifts for the predictions. These are all listed in Table~\ref{tab_arcs}. Future observations of this cluster may reveal a SN in one of the lensed galaxies, in which case these predictions may become useful for interpreting future transients and quick planning of followup. The lens model is also a valuable tool to interpret some of the galaxies in the field. In the next subsections we pay special attention to a few interesting lensed galaxies behind AS1063.

\subsection{A singly imaged but highly magnified galaxy at $z\approx1.85$}
A thin and very elongated arc (system 89 in Table~\ref{tab_arcs}) was already prominent in previous observations with HST in the NE portion of the cluster (Fig.~\ref{Fig_GiantArc}). Earlier attempts to find counterimages for this arc have failed, yet the arc is very stretched (length $\approx 15''$) suggesting that it is strongly lensed and possibly crossing a caustic multiple times. This possibility is reinforced by the proximity of the cluster critical curve, which for $z>2$ crosses the galaxy resulting in multiple images. Careful inspection of the new JWST images reveals a lack of symmetry in the arc, as expected if the galaxy is crossing a caustic ($z\approx 2$), or lack of counterimages that should be easily identifiable in the new JWST images if the galaxy is at ($z>2$). From our lens model we find that the very elongated nature of the arc and lack of symmetry points, or counterimages, can be easily explained if the arc is located in a narrow range of redshift at $z\approx 1.85$. In this situation the arc would lie in a very stretched saddle point in the time delay surface, that is produced by the combined effect of the galaxy cluster and a small group of member galaxies to the NE. The critical curve for $z_s=1.85$ is shown in Fig.~\ref{Fig_GiantArc} as a yellow line.

\begin{figure} 
  \includegraphics[width=9cm]{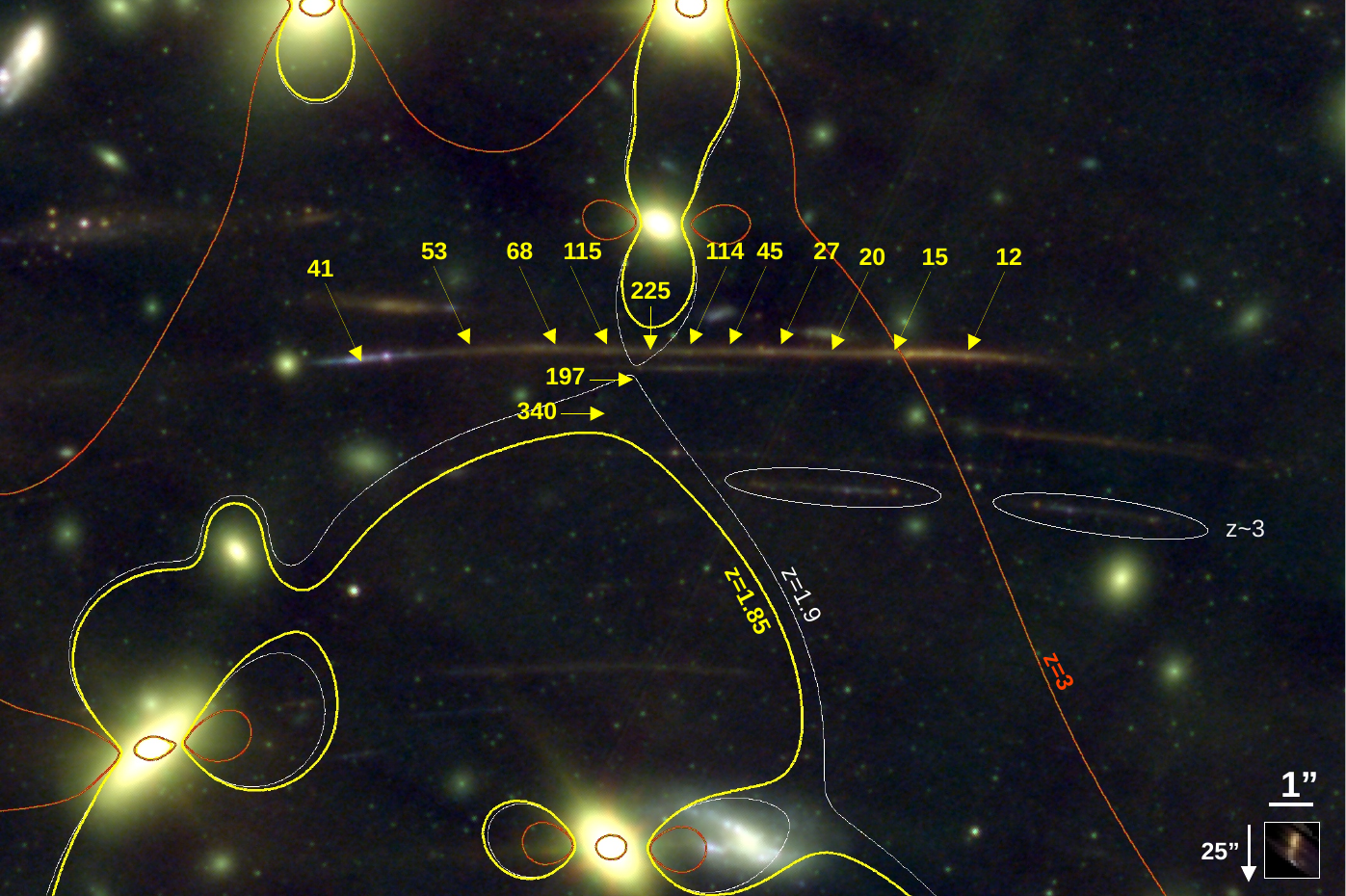}   
      \caption{The giant arc in the north west. Lack of counterimages for this arc and large magnification can be explained only if the arc lies in a narrow redshift range around $z=1.85$. The critical curve for this redshift is shown as a yellow curve, and in white for a slightly larger redshift of $z=1.9$, or red for $z=3$,  where counterimages start to form. The predicted magnification at different points along the giant arc is marked with yellow arrows and numbers. The source plane reconstruction is shown in the bottom-right corner. The entire arc delenses into a merging galaxy system less than 1" in length. South of the giant arc the lens model predicts another unconfirmed system to be at $z\approx3$ (white ellipses) with two symmetric images on each side of the critical curve at $z=3$ (red curve).
         }
         \label{Fig_GiantArc}
\end{figure}

The magnification factor at different points along the arc are also marked with yellow numbers. For reference we also show the critical curves at $z_s=1.9$ (white) and $z_s=3$ (red). In these two cases, the critical curve intersects the arc and produces multiple images, evidence of which can not be found in the data. Instead, if the arc is slightly below $z_s=1.9$, the magnification can be as high as 225 in the middle of the arc and 10 to 40 at the edges. In the bottom-right corner of the figure we show the source reconstruction. This is an interacting galaxy pair of dimension $\approx 4$ kpc, with a blue and a red component, and with its true position $\approx 25"$ towards the cluster center.
If confirmed, this galaxy represents a unique case where the entire galaxy is stretched by a large factor surpassing 100 in the central region and offering a view of the central portion with effective resolutions of better than 1 milliarcsecond for a galaxy at $z\approx 1.85$. In these conditions, microlensing events can be observed almost anywhere along the galaxy. Some of the few unresolved clumps observed in the galaxy could actually correspond to microlensing events which would require a second observation of similar depth to be confirmed, or could be incredibly compact star forming regions or globular clusters in the lensed galaxy (with radii of a few pc at most). 
Also marked in Fig.~\ref{Fig_GiantArc} with white ellipses we show a double image of an arc (containing three knots of the same rank B galaxy, systems 57, 58, and 59 in Table~\ref{tab_arcs}). The geometric redshift  is $\approx 3$ as suggested by the red critical curve at this redshift.

\subsection{A caustic crossing arc at $z=0.73$}
%

\begin{figure} 
  \includegraphics[width=\linewidth]{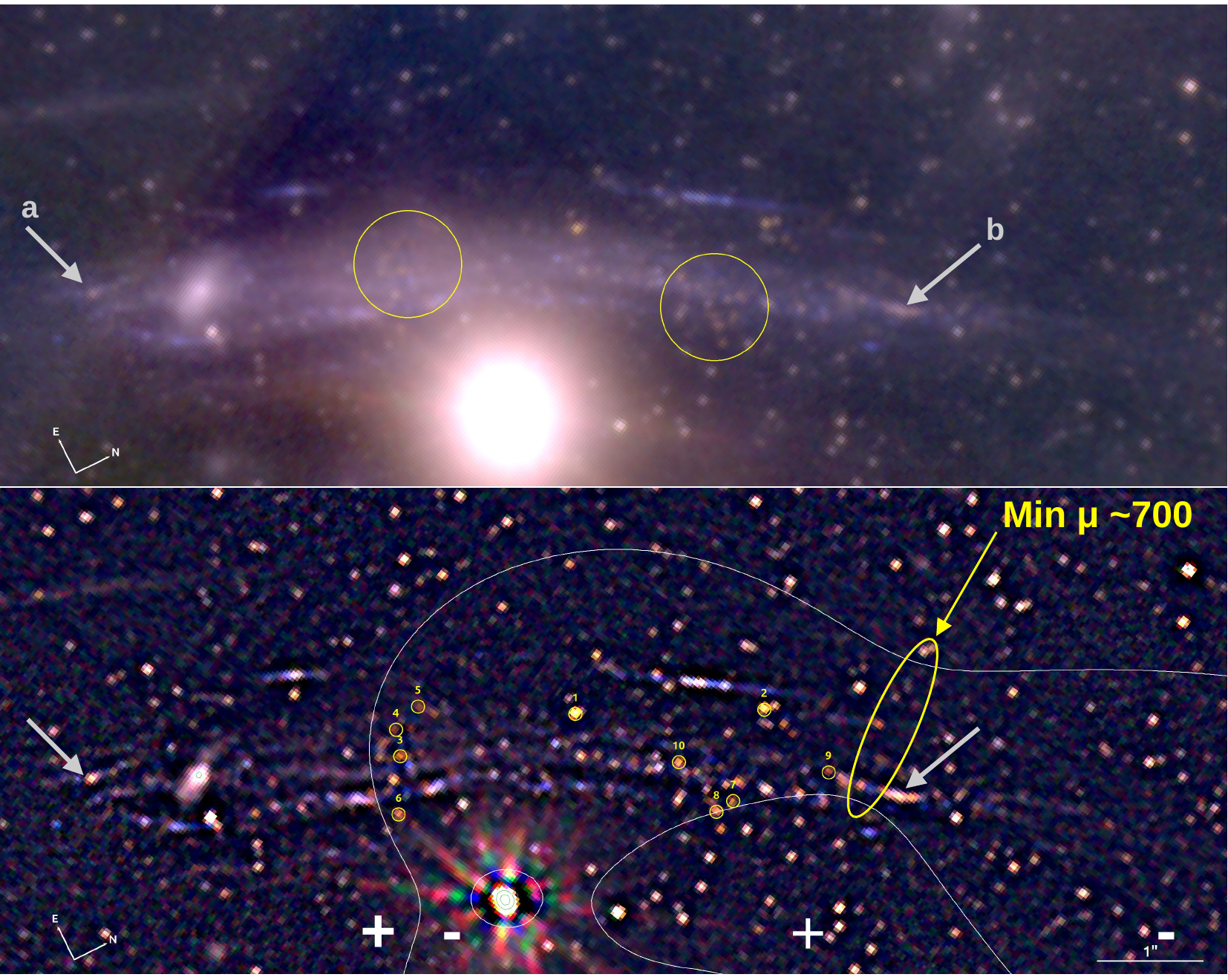} 
      \caption{Color image of the caustic crossing arc at $z=0.73$ (system 19). The top panel shows the color image after combining F090W, F150W and F200W. We mark with arrows a multiply lensed feature in the arc and with two circles regions where we find point sources with distinctive colors (when compared to the surrounding point sources, most of them globular clusters in the galaxy cluster). The bottom image is a color version using a high-pass filter version image of the  bands F090W, F150W and F200W to maximize the color difference of point sources in the area. We show the cluster CC at $z=0.73$ as a white curve and mark with small circles candidate red supergiant lensed stars. The yellow ellipse shows a region where the macromodel magnification alone exceeds 700.  The parity of the images is indicated in the bottom panel with plus and minus symbols. 
         }
         \label{Fig_ArcZ0p73}
\end{figure}

One of the most interesting strongly lensed galaxies in AS1063 is system 19. It was originally identified as a multiply lensed system candidate in \cite{Diego2016b} (their system 10) with geometric redshift $z_{\rm geo}=0.78$ and later confirmed spectroscopically ($z_{\rm spect}=0.73$) by \cite{Caminha2016}. What makes this system interesting is that it is among the lowest redshift caustic crossing galaxies known to date. The redshift is similar to the Dragon arc, that so far holds the record for the largest number of microlensed events detected so far in a single galaxy \citep{Fudamoto25,Palencia2026}. As in the Dragon galaxy, we expect to see many microlensing events in this arc. Two events (likely from two blue supergiant stars) were already reported in this arc with shallower HST data in \cite{Kelly2023}. The new JWST data is much deeper and optimized for the more abundant red supergiant stars, so we would expect to see an even larger number of stars with JWST. Unfortunately, no deep multi-epoch observations exist for this cluster with JWST. However, the deep single-epoch from GLIMPSE reveals tantalizing evidence of individual red giant stars being present as unresolved red sources near the cluster CC. In Fig.~\ref{Fig_ArcZ0p73} we show a color image of the caustic crossing portion of the galaxy. The top panel shows the color image from the SW channels F090W, F150W and F200W while the bottom panel is a different color image based on the same filters, but after removing the largest scales with a high-pass filter. Most of the unresolved point sources in the image are GCs in AS1063 or small unresolved stellar groups in the arc itself, but at the position of the lensed galaxy we see point sources which appear more red. Especially near the CC where magnification is higher, and the probability of red supergiants to experience microlensing is greatest. 
In the lower panel of  Fig.~\ref{Fig_ArcZ0p73} we mark with circles some candidate red supergiant stars. These are distinctively more red than most of the surrounding point sources. We find these candidates near the region of highest magnification, as expected for microlensing events. The photometry of these sources is shown in Fig.~\ref{Fig_ArcZ0p73_Photometry}. Magnitudes are computed in apertures of 0".09, with background and uncertainty computed in similar apertures around each target. The photometry it is compared to the synthetic spectrum \cite[from][]{Coelho2014} of a typical red supergiant star with temperature 4000 K. All sources peak at the redshifted 1.6\, $\mu$m bump assuming they are at the redshift of the arc ($z=0.73$) this supporting the interpretation that these are point sources in the lensed galaxy.  Sources 1 and 2 are in the outskirts of the arc but still consistent with being highly magnified supergiants.  Source 1 is particularly interesting because it is relatively farther away from the critical curve, yet it appears as the brightest source among the candidates. The apparent magnitude of sources 1 and 2 are consistent with the typical absolute magnitude in H band of a red supergiant, -9, at redshift $z=0.73$ (or distance modulus 43.26) and at relatively moderate magnification factors of a few hundred. At larger magnification factors typical of microlensing events (few thousand), fainter AGB stars can be seen at these magnitudes as well. Even stars at the tip of the red giant branch (TRGB) should be detectable in GLIMPSE data during microlensing events at $z=0.73$. At absolute magnitude in H band $\approx -6$ \citep{Newman2024}, and after transforming Vega to AB magnitudes \citep[+1.3618 in $I$ band, see Table 4 in][]{MaizApellaniz2007}, and band correction, $-2.5{\rm log}_{10}(1+z)$), it takes magnification $\mu\approx3000$ for a star in the TRGB to appear as a source with apparent magnitude $\approx 29.3$ AB in F277W (shown in Fig.~\ref{Fig_ArcZ0p73_Photometry}).  These magnification factors are typical near the peak of microlensing events of red supergiant stars, and can be maintained for a few weeks \citep{Diego2026c}. This opens the interesting possibility of using the J-AGB and TRGB methods to derive an independent distance estimator to this galaxy \citep[see the detailed case study for the Dragon arc at a similar redshift in][]{Diego2026a}.

\begin{figure} 
  \includegraphics[width=\linewidth]{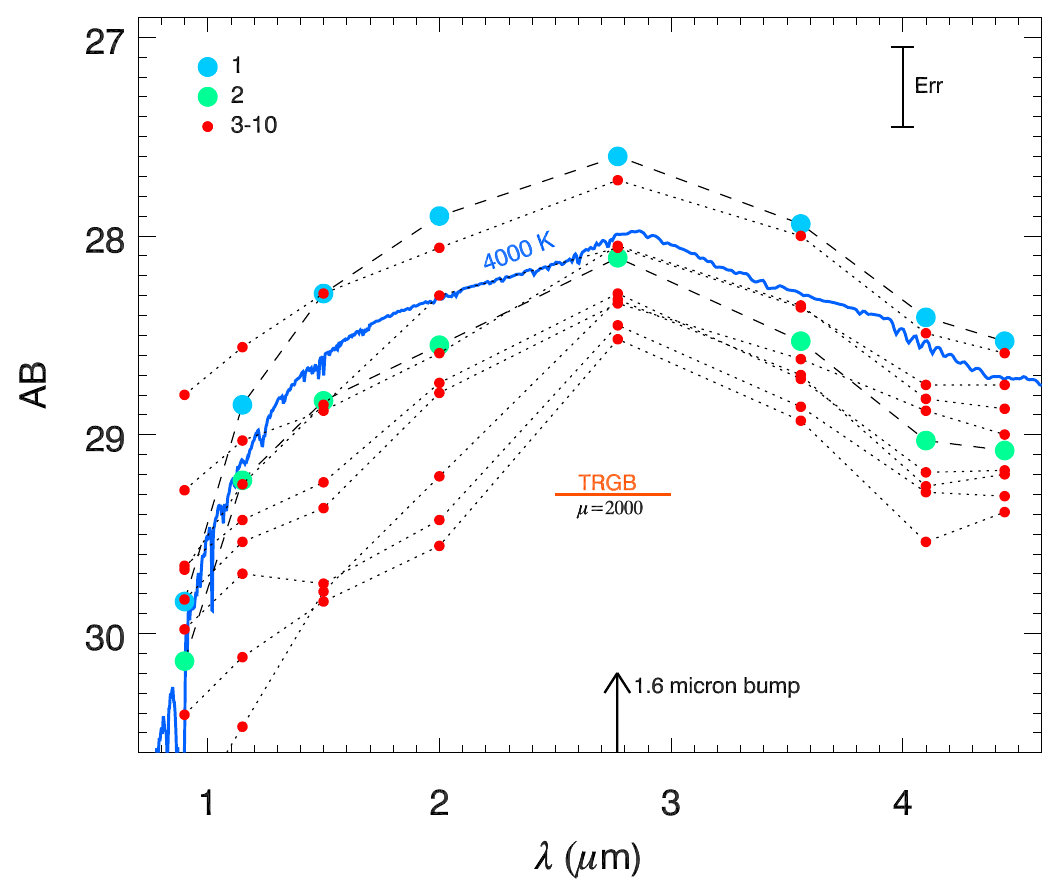} 
      \caption{Photometry of the red supergiant candidates at $z=0.73$ shown in Fig.~\ref{Fig_ArcZ0p73}. Numbers 1 to 10 correspond to the circles with the same labels in Fig.~\ref{Fig_ArcZ0p73}. Sources 1 and 2 are marked with different symbols to distinguish them from other sources. For simplicity, photometric errors are not included but a typical error is shown in the top-right corner. For reference, source 1 is at RA=342.1896392, DEC=-44.5291789. The expected lensed TRGB is show at AB 29.8.
         }
         \label{Fig_ArcZ0p73_Photometry}
\end{figure}
If these red sources are indeed highly magnified stars, future deep observations of AS1063 with JWST should reveal changes in the flux of most of these red objects, since changes in magnification of lensed stars due to microlensing should be frequent at this short distances from the cluster center and proximity to the CC. The lens model also reveals a region of extremely high macromodel magnification found between two CCs (marked with a yellow ellipse), having macromodel magnification of at least 700. Only a small portion of the caustic crossing galaxy intersects this region, but the large magnification provides a unique opportunity to study this region of the galaxy at $z=0.73$ as if it were $\sqrt{700}=26.5$ times closer, or $z=0.04$, and without being disrupted by a caustic (since it is between two caustics). We find one multiply lensed structure in this region of high magnification, labeled "a" and "b" in the figure. While "a" is still unresolved, "b" is much more stretched consistent with the much higher magnification predicted by our model.

\subsection{EELG or LRD}\label{sec_LRD}

\begin{figure} 
  \includegraphics[width=\linewidth]{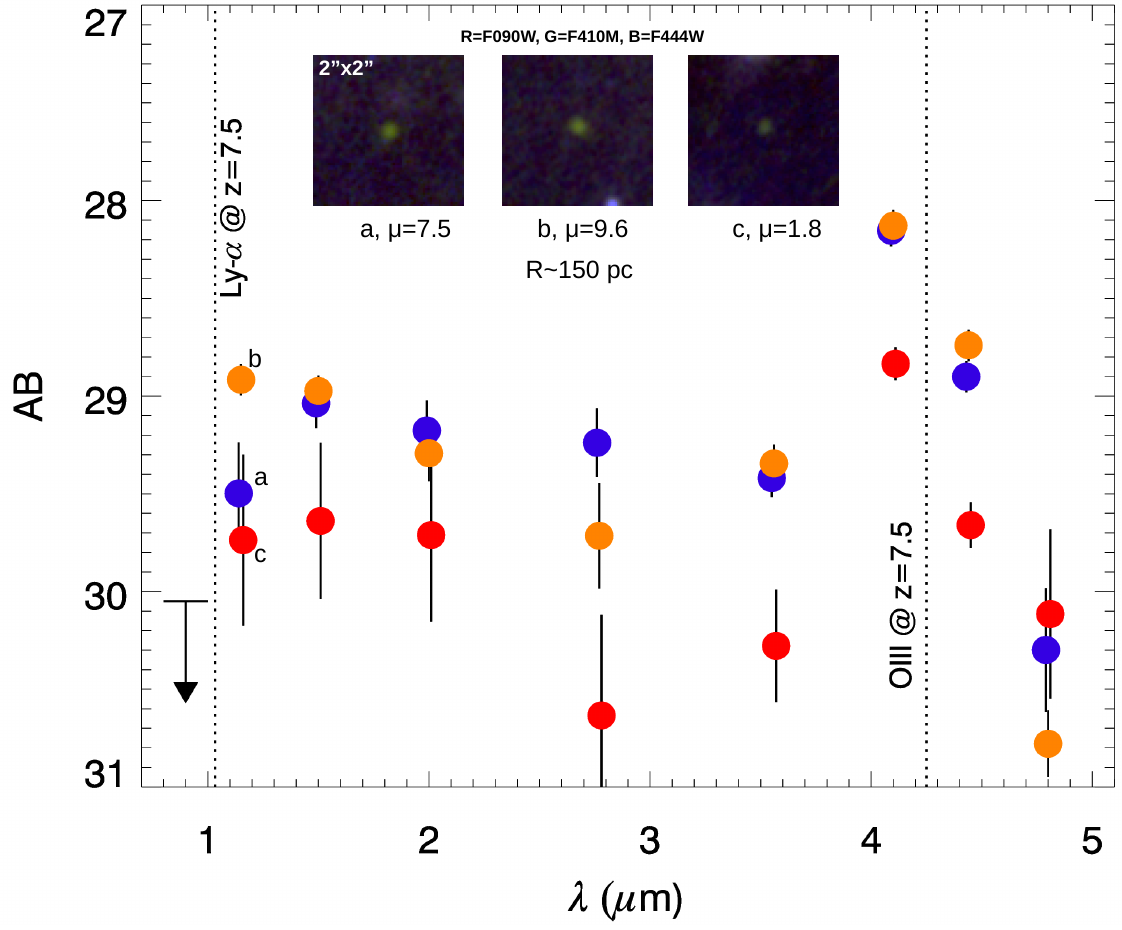} 
      \caption{Multiply lensed ELG (new system 78) found after searching for high-z lensed galaxies near the $z=10$ CC. The galaxy is not detected in F090W and shows prominent emission in F410M. The three counterimages are consistent in color an predicted position if $z\approx 7.5$. The predicted magnification at this redshift is also consistent with the flux ratio specially if one accounts for the possibly biased low magnification of image c. At this redshift, the emission observed in F410M would correspond to H$\beta$--OIII. 
         }
         \label{Fig_ELG}
\end{figure}

Among the new high-redshift lensed galaxies, we have discovered one that is multiply imaged into three counterimages (system 78 in Table~\ref{tab_arcs}), all of them being dropouts in F090W and with prominent emission in F410M. Both the dropout nature and strong emission at $\approx 4\, \mu$m are consistent with a galaxy at $z\approx 7.5$ and strong H$\beta$--OIII emission. We show the photometry in Fig.~\ref{Fig_ELG}. 
Magnitudes are computed in circular apertures of radius 0."2. This radius ensures that in the widest filters most of the flux is contained within this aperture. Background an uncertainty is computed using 3 control apertures near and around the main target for each of the three positions, making sure no other source near the main targets fall within the control apertures. We use the control apertures to estimate the background (mean) and uncertainty (dispersion).
The restframe UV part of the spectrum and strong emission at 4 $\mu$m suggests that this is an extreme emission line galaxy (EELG) or given its very compact nature a little red dot (LRD). The image 78.b is the most magnified and its is barely resolved in the SW JWST channels, implying a physical size of $R\sim 150$ pc (radius). We detect no sign of the Balmer break expected for LRD but this is consistent with having a very large equivalent width for the OIII line \citep[see Fig. 14 in][]{deGraaff2025b}. The predicted time delay between images is as long as 78.6 years allowing to check for long duration intrinsic variability by comparing the fluxes of different images with the magnification from the lens model \citep{Zhang2025}. The time delay between images "a" and "b" is 4.1 months, or two weeks in the rest frame, with photons in image "b" arriving before photons in image "a". 

\section{Hedorah, a yellow supergiant monster star candidate at $z= 3.72$}\label{sec_Hedorah}
The most intriguing object we find in this cluster is an unresolved source in a caustic crossing galaxy (system 60 in Table~\ref{tab_arcs}) and very close to the critical curve ($d=0."18$). It shows a very peculiar yellowish color (in the rest frame) that does not resemble any of the point sources nearby, including the GCs. Among the possibilities explored, we conclude in section~\ref{sec_discussion} that a strongly lensed monster star is the most plausible candidate. Because of its yellow color, we nickname it ''Hedorah'', following the tradition of previously discovered highly magnified monster stars, and honoring the yellow-eyed Kaiju. \\

\begin{figure} 
  \includegraphics[width=\linewidth]{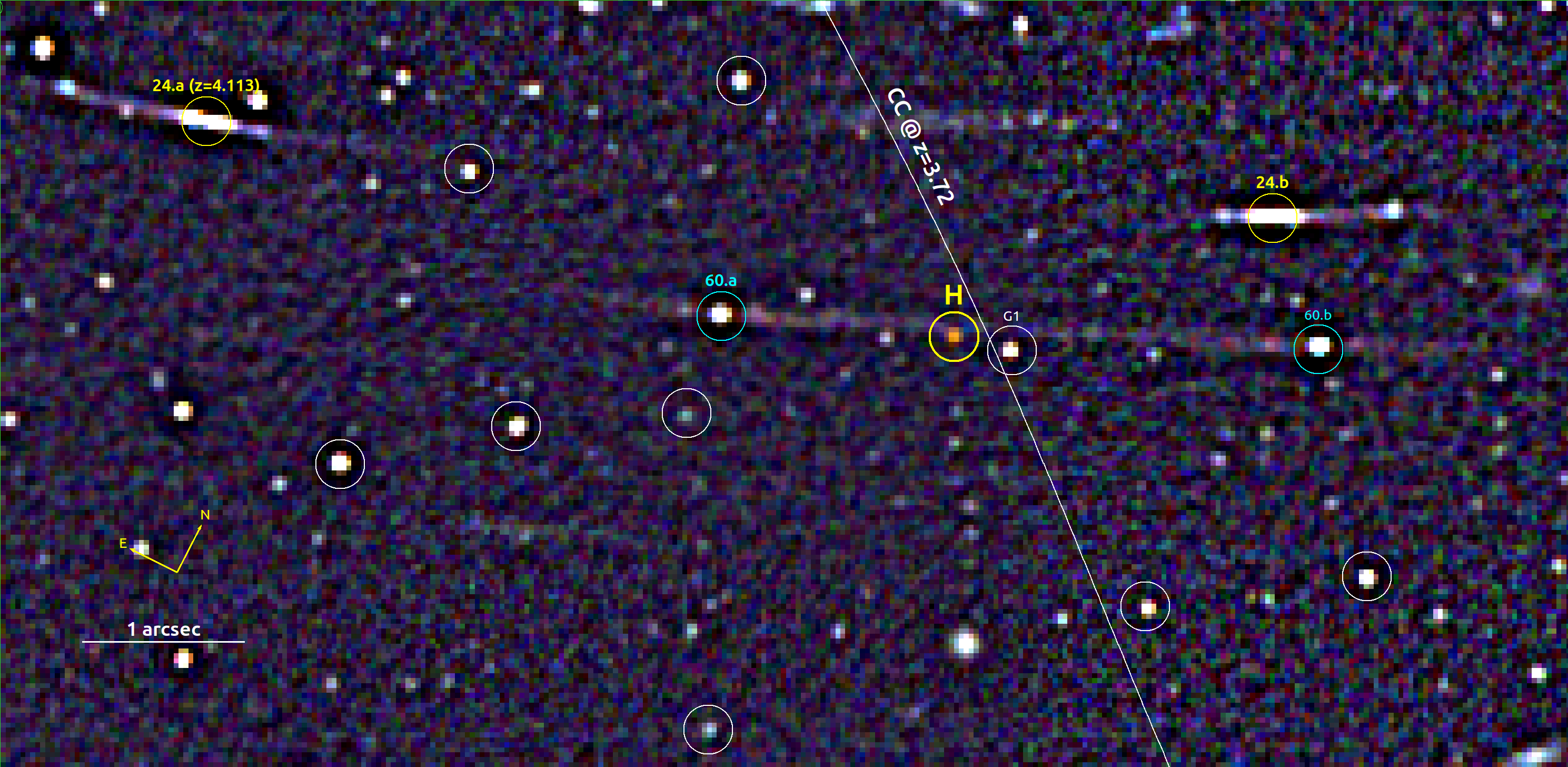}   
      \caption{Hedorah (RA=342.1796997, DEC=-44.5391334), marked with an H, a candidate microlensing event of a lensed yellow supergiant star at $z\approx 3.7$. The color image shows the combination of the high-pass filtered version of the RGB image F090W+F150W+F200W. The white line marks the CC at $z=3.72$.  The green circle labeled G1 is a possible clump in the lensed galaxy, but more consistent with being a GC in AS1063. The remaining green circles show eight randomly selected GCs used to derive their photometry and compare it with Hedorah's (see Fig.~\ref{Fig_SED_Kaiju})
         }
         \label{Fig_RSG1}
\end{figure}

After the original submission of this work, \cite{Furtak2026} measured the spectroscopic redshift of its host galaxy, $z=3.72$, or $20\%$ higher than the (predicted) geometric redshift predicted by the lens model, $z_{\rm geo}\approx 3.1\pm0.2$, (system 60 in Table A.1). 
This source is undetected in F090W and barely detected in F115W. It shows significant emission in F150W and relatively flat emission between F200W and F480M (see Fig.~\ref{Fig_SED_Kaiju}). Its photometry is consistent with that of a single star at the redshift of the arc ($z\approx 3.7$) and temperature $T=6500$ K (Fig.~\ref{Fig_SED_Kaiju}), that is a yellow supergiant star or yellow hypergiant star. 
The photometry was done using circular apertures of radius $0."15$,  with the background and uncertainty obtained from similar apertures distributed around the main target (and not overlapping with it).

If Hedorah were one of the GCs in the cluster at $z=0.348$ we would expect a nearly constant magnitude between F115W and F277W and a small decline in F090W. This is observed in the nearby bright clumps found near Hedorah, ruling out the GC interpretation (see inset in Fig.~\ref{Fig_SED_Kaiju}, where we use the same aperture radius and procedure described above to compute the magnitudes of the GCs.). The GCs (black dots) show the expected bump at $1.6\, \mu$m for an aged stellar population at $z=0.348$ redshifted into F200W \citep{Sawicki2002}. The inset in Fig.~\ref{Fig_SED_Kaiju} shows how the clump G1 is also consistent with the SED of the GCs. Together with the fact that this clumps has no clear counterimage (on the side of positive parity where demagnification is not possible), makes this clump likely a compact object in the lens. GCs act as mililenses, with lensing distortions extending only  miliarcseconds from each GC. All visible GCs, including G1, are at least 0."2 from Hedorah and hence have negligible impact regarding lensing distortions on Hedorah.  However, G1 sits close to the expected position of Hedorah's counterimage and can demagnify it. This could explain the lack of detection of Hedorah's counterimage (see next section). \\

A second possibility is that Hedorah is a clump of an unrelated source at a different redshift. As long as the redshift of this source is $z>1$ there should be counterimages at predictable positions that are not found at the lens model predicted positions. Despite being an unresolved source, the color is very unique with no other unresolved sources matching this peculiar color at any of the predicted positions for $z>1$. For lower redshifts, the situation is similar to the GC in the lens. The dropout nature in F090W and very steep rise between $1.15\, \mu$m and $2\, \mu$m would require a source with extreme reddening, a quality that is not supported by the flat spectrum shown in Fig.~\ref{Fig_SED_Kaiju} above $2\, \mu$m. A third possibility is that Hedorah is not an extremely magnified lensed star but a still magnified (by a more modest factor), small clump of stars in the host galaxy at $z\approx3.7$. Based on the lens model, and ignoring small scale perturbations in the magnification from microlenses or millilenses, the predicted macromodel magnification at the position of Hedorah  is $\mu\approx 360$.  

\begin{figure} 
  \includegraphics[width=\linewidth]{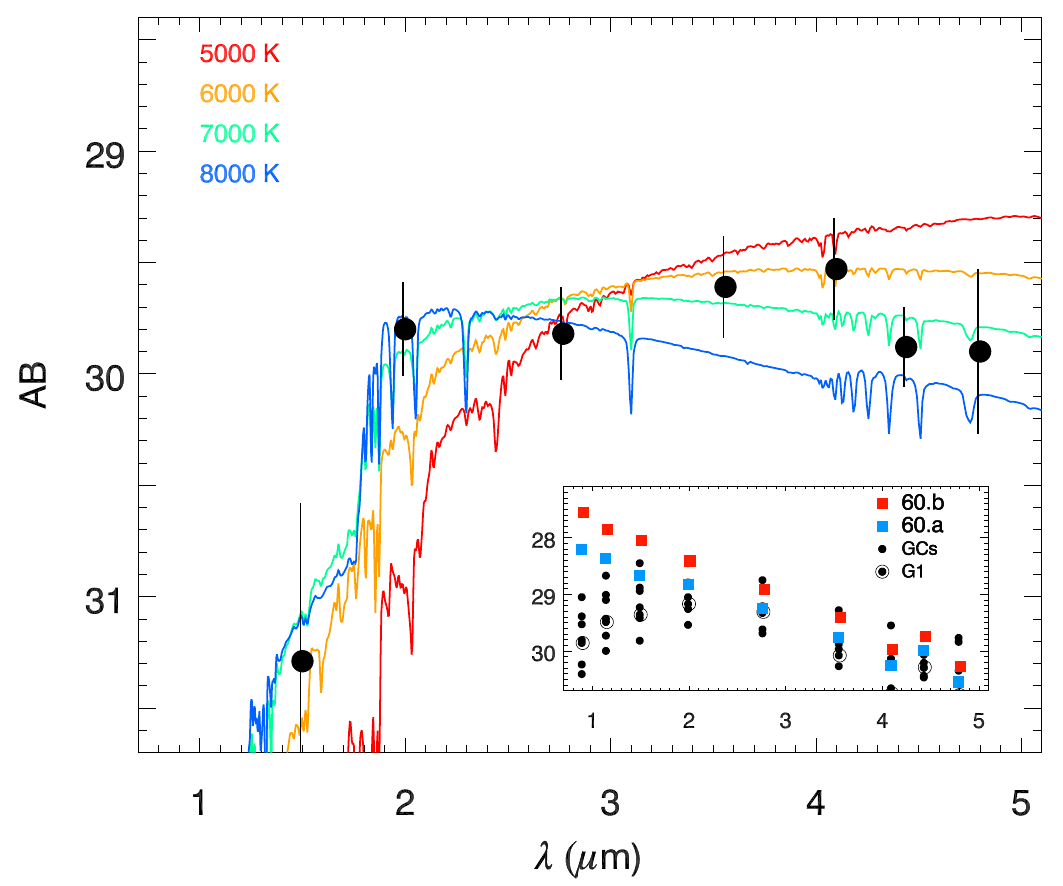} 
      \caption{SED of the lensed star candidate Hedorah. Four stellar models with different temperatures and at the spectroscopic redshift of the host galaxy, $z=3.72$, are shown \citep[from][]{Coelho2014}. The data is consistent with a yellow supergiant star with temperature $\approx 6500$ K at $z=3.72$. The inset shows the SED of unresolved GCs nearby (including G1) and the bright clumps 60.a and 60.b. 
         }
         \label{Fig_SED_Kaiju}
\end{figure}

Hedorah is close to the cluster CC ($d=0\farcs18$) but not exactly on the CC, hence a counterimage with similar color and flux is expected at a similar distance from the CC but on the opposite side. 
However, no counterimage for Hedorah is found on the other side of the CC, even in the ultradeep stacked image (HST+JWST). One possibility is that a substructure is demagnifying the counterimage making it undetectable. This is certainly possible because the counterimage is on the side of the critical curve where saddle images (or images with negative parity) form. Images with negative parity can be demagnified by microlenses and millilenses. If Hedorah is a small compact group of stars, a microlens is not massive enough to demagnify an entire group of stars so a more massive millilens is needed. No sign of the millilens can be found near the predicted position of the counterimage but this does not rule out the possibility of a small millilens exiting at that position. We explore this possibility in more detail in Sect.~\ref{sec_discussion}. 
 The final alternative scenario is a single star undergoing a microlensing event. This possibility requires fewer ingredients and naturally explains the lack of counterimage since the  microlensing event would take place in one of the counterimages (Hedorah) but not in the other. In addition, yellow supergiants are typically variable stars (for instance Cepheids stars) naturally explaining the flux difference between counterimages. We also explore this scenario in more detail in Sect.~\ref{sec_discussion}. 

\section{The nature of Hedorah} \label{sec_discussion}

Having ruled out the possibility that Hedorah is a GC in the lens, or a small galaxy below $z=3$, the  most likely interpretations for Hedorah can be narrowed down to i) a clump with very peculiar color in the lensed arc at $z\approx3.7$, with a counterimage hidden by an invisible substructure, or ii) a yellow supergiant or hypergiant star observed during a maximum in brightness due to intrinsic variability (for instance a Cepheid star) or a microlensing event. Here we explore these two scenarios in more detail.

 {\bf i) Star clump and dark substructure:} This scenario requires two exotic components, a clump with very yellowish color matching well the spectrum of a single star with $T\approx 6500$ K and a dark or invisible millilens at the position of the counterimage. The invisible millilens is not that exotic since it could simply be one of the many GCs that is less luminous than the ones observed in JWST's images, but this millilens needs to be located precisely in front of the counterimage to demagnify it. To assess the probability that a millilens can be demagnifying Hedorah's counterimage, we estimated the local number density of GCs around Hedorah's counterimage. We applied a high-pass filter to the JWST images in the SW channels to optimize the detection of GCs and found a number density of $\approx 2.3\pm 0.4$ GCs per arcsec$^2$ around the arc hosting Hedorah (computed in boxes of $5"\times5"$). No point source is found near the position of the expected counterimage of Hedorah, even in the ultradeep image obtained after stacking the data taken in the HST and JWST filters (about one magnitude deeper). Hence, if the alleged millilens is an undetected GC, its mass must be relatively small ($M<10^5\, \Msun$) to escape detection by JWST. At this mass, the area of influence of a millilens is very small. Even after accounting for the magnification from the macromodel (that amplifies also the effective lensing mass of the millilens) this area remains small. As an example we show in Fig.~\ref{Fig_Millilens} a $10^4\, \Msun$ millilens at the positions of Hedorah (positive parity, left) and its counterimage (negative parity, right). At these positions, the macromodel magnification takes the value 
$\mu_{\rm macro}=\mu_t\times\mu_r=\pm225\times1.6)\pm360$. 
A $10^4\, \Msun$ millilens in front of Hedorah can certainly amplify its flux by a factor $\lesssim 2$ if Hedorah is a small group of radius $r\approx 0.3$ pc. If the same millilens is placed right along the line of sight to Hedorah's counterimage, it can completely demagnify it as long as Hedorah is a group with radius $r\lesssim0.3$ pc and it is perfectly aligned with the millilens. The region where demagnification can take place is $\approx 0.1\times1$ mas$^2$, or $\approx 0.7\times7.2$ pc$^2$ at $z\approx3.7$. However, the probability of this perfect alignment is very small given the number density of $\approx 2.3\pm 0.4$ GCs per arcsec$^2$ found in the images. Even assuming a factor ten more millilenses, or $20$ per arcsec$^2$ (to include the ones falling below the detection capabilities of the deep JWST images), and another factor ten for the mass of the millilenses, the probability of one of this GCs to fall in the required 0.1 mas$^2$ region is incredibly small $P=2\times10^{-5}$. The alignment does not need to be perfect since the demagnification does not need to be that pronounced, so we can relax this condition by a factor $\approx 5$ which increases the probability to $P\approx 10^{-4}$. Although not impossible, this scenario is highly unlikely. Furthermore, in addition to the small probability of Hedorah's counterimage being demagnified by a small enough millilens, the small star group explaining Hedorah's observed flux still needs to be dominated by stars with a spectrum that resembles that of yellow supergiants as shown in Fig.~\ref{Fig_SED_Kaiju}. Yellow supergiant stars or yellow hypergiant stars are exceedingly rare, so the mix of stars making up Hedorah's group needs to be fine tuned between older red stars, and younger blue stars so the combined spectra resembles the observed one. However, since there is no sign of blue emission from redshifted hot stars in the observed spectrum, the star group should also include large amounts of dust that absorb the UV emission. There is no sign of this dust in the two bright clumps of the host galaxy, 60.a and 60.b, that show a very blue and unabsorbed spectrum (see inset in Fig.~\ref{Fig_SED_Kaiju}). At even longer wavelengths, if Hedorah were to be an aged GC at $z\approx3.7$ we would expect the color F356W-F444W to be slightly positive since the $1.6\, \mu$m bump redshifts into 7.5 $\mu$m \cite{Muzzin2013}, while we find a negative value (Fig.~\ref{Fig_SED_Kaiju}) which would imply a very young age ($<10$ Myr). Hence we conclude that scenario i) is unlikely based on the low probability of a small millilens being perfectly aligned with the counterimage and the unique spectrum and colors of Hedorah. \\

\begin{figure} 
  \includegraphics[width=\linewidth]{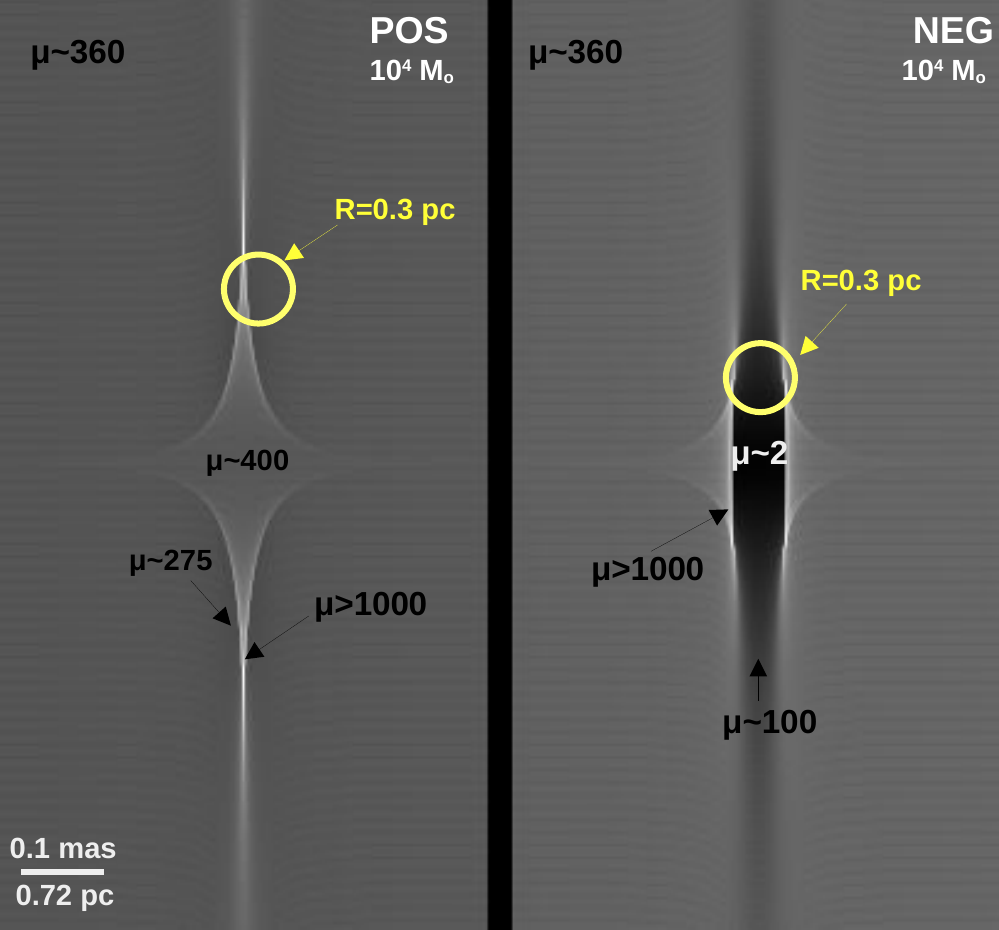} 
      \caption{Caustics of a compact $10^4\, {\rm M}_{\odot}$ millilens in the lens near the CC  with macromodel magnification $\mu=360$. For presentation purposes, the caustic region has been compressed a factor 8 in the vertical direction which spans 9 mas.  A small $R=0.3$ pc cluster of stars is shown (in projection) by the millilens caustic on the side of the CC with positive parity (left) and negative parity (right). Typical magnification factors, $\mu$, are marked. 
         }
         \label{Fig_Millilens}
\end{figure}

{\bf ii) A yellow supergiant or hypergiant star :} 
This scenario is simpler because it requires fewer exotic components. Yellow supergiant stars (although rare) and numerous microlenses are both expected to exist in the arc and in the lens respectively. Not observing the counterimage is easier to explain if Hedorah is experiencing microlensing or it is a variable star (as most yellow stars) observed during a maximum in one image but not the other. However, the possibility that Hedorah's variability explains the lack of detection of its counterimage is problematic. The predicted time delay between Hedorah and its counterimage is $\approx 2$ weeks. Accounting for time dilation, this requires that the period of Hedorah must be less than one week. If Hedorah is a variable star with a  regular period, for instance a Cepheid, short period Cepheids ($P<1$ week) are typically not luminous enough (owing to the period-luminosity relation of Cepheids), even after magnification by a microlens, to be observed at this redshift. If Hedorah is a Cepheid at $z\approx3.7$ with a  period of less than a week, it would require significantly more magnification than what can be provided by microlenses ($\mu<10000$) and would require a more massive millilens. This possibility suffers of the same issues discussed in scenario i) above, namely the very small probability of having a small millilens at Hedorah's position. Hence, we ignore short period Cepheids and consider only the case of a Cepheid with a long period experiencing microlensing. 
The absolute (Vega) magnitude  of Cepheids in the $I$-band relates to their period as: 
\begin{equation}
M_I = -2.81[{\log}_{10}(P)-1.0] -4.76,
\end{equation}
where $P$ is expressed in days \citep{Storm2011}. 
 At $z=3.72$, the $I$-band redshifts into the F356W filters, where Hedorah is detected by JWST at AB$\approx 29.6$. If we consider a very long period (and luminous) Cepheid with $P=100$ days, or $M_I=-7.6$, this becomes AB -8.86 mags after transforming Vega to AB magnitudes \citep[+0.419 in $I$ band, see Table 4 in][]{MaizApellaniz2007}, and band correction, $-2.5{\rm log}_{10}(1+z)$). 

\begin{figure} 
  \includegraphics[width=\linewidth]{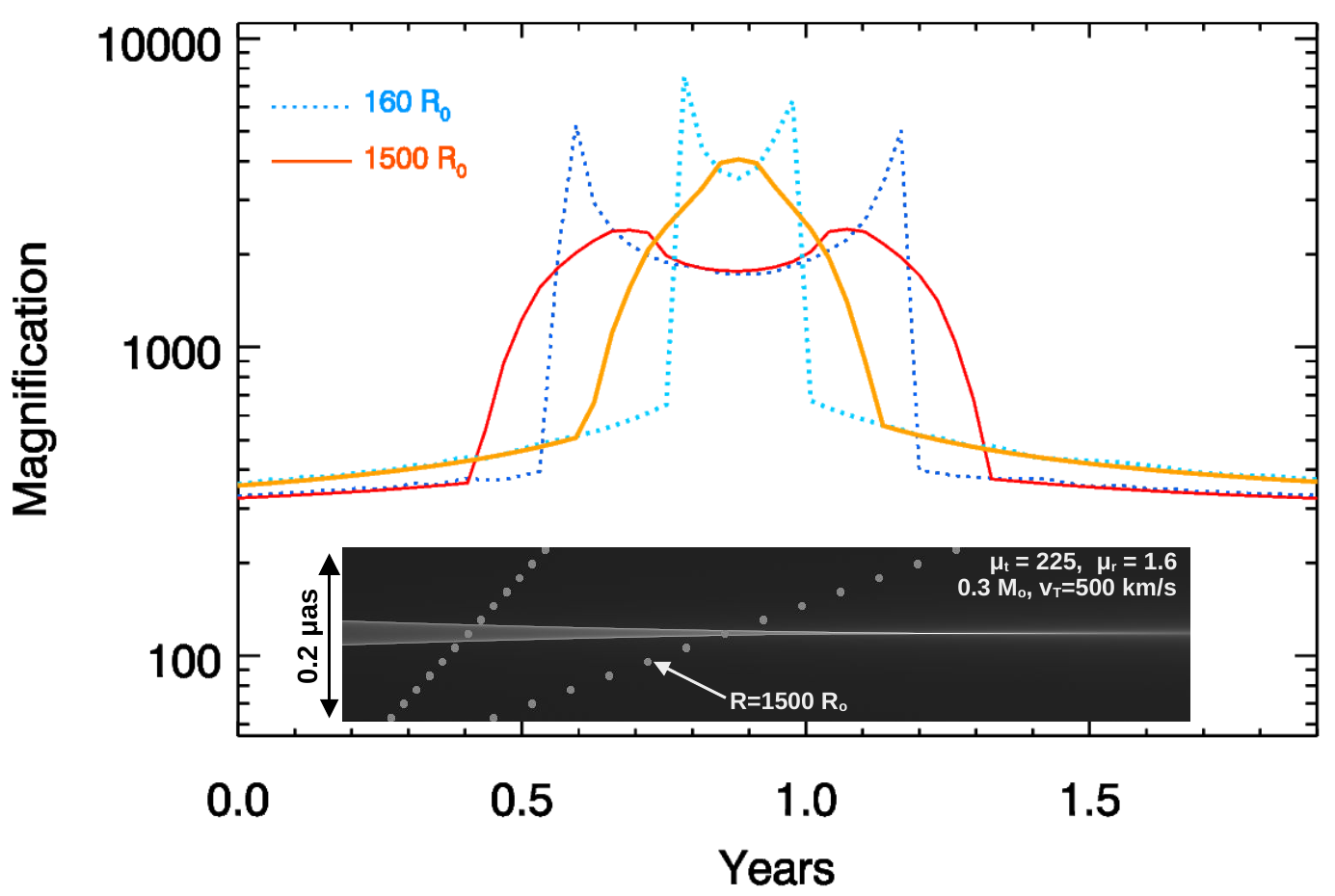} 
      \caption{Lightcurves of stars crossing a cusp region of a $0.3\, \Msun$ microlens in the galaxy cluster and along the line of sight to Hedorah. The bottom panel shows a portion of the caustic region with tracks of two stars of radii $R=1500\, \Rsun$. The corresponding light curves are shown in the main panel in red (and orange). In both cases, the tangential component of the velocity of both stars is 500 km s$^{-1}$. For comparison we show in blue the magnification of a smaller star following the same trajectories but with diameter equal to the pixel size of the simulation (1 nas or $R=160\, \Rsun$). The small scale fluctuations are an artifact from the inverse ray shooting. 
         }
         \label{Fig_Microlens}
\end{figure}

At distance modulus 47.6 ($z=3.7$), such star would have apparent magnitude AB 38.9. In order to be detected by JWST in F356W at AB 29.6 mags, the Cepheid needs to be magnified by a factor $\mu=5000$. This is in principle possible by the combined effect of the macromodel and one of the abundant microlenses along the line of sight to Hedorah, specially one from the intracluster medium (ICM). Stars in the ICM are typically not very massive since they are relatively old stars.  In Fig.~\ref{Fig_Microlens} we show a microlensing event for a typical microlens in the intracluster medium ($0.3\, \Msun$) and two possible paths for the Hedorah star. The solid red (and orange) curves correspond to a star with a very large radius of $r=1500\, \Rsun$ with the two paths shown in the lower portion of the figure as they cross the microcaustic. The blue dotted lines are for a smaller star following the same paths, but with diameter equal to the pixel size of the simulation, or $160\, \Rsun$ radius, similar to that of long period Cepheids. In all cases, the star is moving with the tangential component of the velocity (i.e., perpendicular to the microcaustic) equal to 500 km s$^{-1}$. This is the typical expected relative velocity for the redshifts involved in this problem (observer, lens and source planes are all moving relative to each other). The larger star (red curves) is magnified less at the peaks owing to the maximum magnification scaling as $\mu_{\rm max}\propto 1/\sqrt{\Rsun}$. 
For this particular case, and focusing on the blue curves which are more representative of long period Cepheids, magnification factors $\mu\approx 5000$ can be obtained at the caustics during short periods ($\sim$ 2-4 weeks).  Magnification factors $\mu>5000$ are only possible in this configuration in the valley between caustics if the star is really close to the tip of the cusp, in which case these high magnification values can be sustained for about a month. Larger magnification values could be achieved with the microlens shown in Fig.~\ref{Fig_Microlens} at the caustics if Hedorah's radius is less than $160\, \Rsun$ or if the microlens is more massive. More massive microlenses are expected in the intracluster medium, for instance a remnant (neutron star or black hole), or more exotic microlenses such as primordial black holes with even larger masses.  But these massive microlenses are more rare than the more common stars responsible for the intracluster light, so the possibility that Hedorah is a long period classic Cepheid is relatively low, as it requires incredible luck to have been observed during the JWST observations, or it needs a more complex lensing configuration (not explored here and beyond the scope of the paper). For instance the existence of a millilens nearby further boosting the magnification from the microlens, or more than one microlens working together increasing again the combined magnification. In both cases Hedorah can remain above the detection threshold for longer periods. There is also the unrealistic case where the relative velocity between Hedorah and a microcaustic is much smaller than the one considered here. A relative velocity of just 10 km s$^{-1}$ would allow Hedorah to remain near the microcaustic of a massive microlens (or microlens binary) and above magnification $\mu=2000$ for several months, but this is also highly unlikely so one needs to find a better explanation. 
The macromodel magnification can also be smaller than the value considered here. For instance, in a recent work, \cite{Furtak2026} estimates a macromodel magnification for Hedorah $\mu=98$, or about 3.5 times smaller than our estimation. If confirmed, it would make the Cepheid interpretation of Hedorah even more challenging.

Hedorah is still likely a star, but not necessarily a classic Cepheid. Extraordinarily luminous stars would more easily remain above the detection threshold for longer periods. Yellow supergiants can be about 1 magnitude brighter than the long period Cepheid considered above. They are very rare but they do exist, perhaps in larger numbers at $z\approx 3$. The more luminous a star is, the more likely it is to be found farther away from CC. If, as predicted by the alternative model of \cite{Furtak2026} ($d_{CC}=0."62)$, the separation from Hedorah to the CC is significantly larger than the $d_{CC}\approx 0."18$ predicted by our lens model, Hedorah would be among the farthest stars from a CC ever found, specially if we ignore the stars detected in the Dragon Arc \citep[$z=0.725$,][]{Fudamoto25}, which owing to their smallest redshift allows intrinsically fainter stars to be detected at smaller magnifications (and hence farther from the CC). Specially if we consider the fact that Hedorah is at $z=3.72$, and high redshift stars tend to be observed very close to the CC. We can then assume that Hedorah is a true monster star with incredible luminosity. Yellow hypergiants, such as  IRC+10420, show similarities with Hedorah and can serve as local analogs. IRC+10420 shows resemblance also with another Milky way monster star, $\eta$-Carinae \cite{Neugebauer1969,Humphreys1973} and it is classified as a variable F Ia+ class star (similar temperature in the past to Hedorah, but currently experiencing an increase in its temperature as it evolves into a blue supergiant star).  IRC+10420 shows strong H$\alpha$ lines and prominent photospheric lines with strong wind features \cite{Thompson1977,Jones1993,Humphreys2002}. Intriguingly, its spectrum shows two bumps in the continuum at around 1 $\mu$m and 1.25 $\mu$m \citep{Oudmaijer1996}. When redshifted, the first bump corresponds nicely with the increase in flux observed in Hedorah at $\approx 4\, \mu$m, while the second one is outside the range covered by JWST observations.
IRC+10420 has been interpreted as a star evolving from red to blue supergiant, and possibly ending its life as a Wolf-Rayet star \citep{Jones1993,Humphreys1997}. IRC+10420 is believed to have currently halted its evolution at the top of the HR diagram and about to experience a major disruptive event \citep{Koumpia2022}, suggested as well by its strong maser emission \citep{Giguere1976,Reid1979}. The idea that Hedorah is a star at $z\approx3.7$ approaching its Wolf-Rayet phase is tantalizing. Hedorah could represent the most distant example of a possible SN progenitor, which so far have been studied only in the local universe \citep{VanDyk2026}. 
At AB$\approx 29$ it represents a formidable and challenging target for spectroscopy, possibly out of reach for telescopes such as JWST, but perhaps within the capabilities of soon-to-be-reality new generation telescopes such as the ELT.   
Other interesting analogs are V382 Carinae, a pseudo-Cepheid star, or the yellow hypergiant Rho Cassiopeiae with  M$_{\rm v} \approx -9.5$ and temperature $6500<T_{\rm eff}<7200$ \citep{Percy2000,Gorlova2006}.\\

\begin{figure} 
  \includegraphics[width=\linewidth]{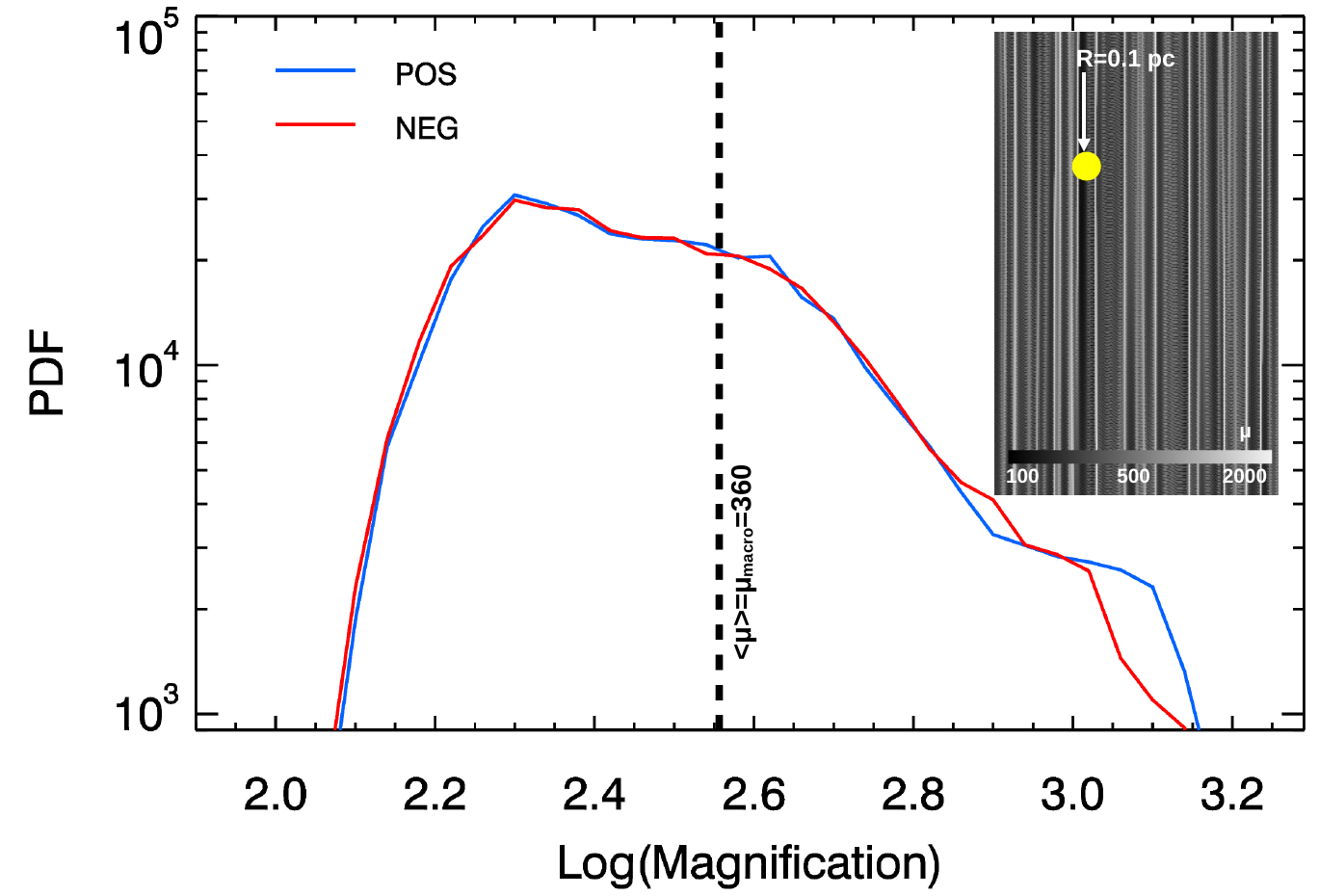}    
      \caption{Probability of magnification for $\psi$DM at the position of Hedorah (POS) and its counterimage (NEG). The inset shows the magnification in the source plane at the position of Hedorah's counterimage. For reference, a small circular structure with radius 0.1 pc is shown as a yellow circle. For this model $\lambda_{\rm dB}=10$ pc.
      The vertical dashed line marks the macromodel magnification which coincides with the average value of the simulation after adding the $\psi$DM fluctuations. }
         \label{Fig_PDF_WaveDM}
\end{figure}

\noindent
Alternative dark matter scenarios : If the dark matter particle is ultralight, with masses well below the QCD axion, interesting lensing effects appear \citep{Schive2014a,Schive2014b,Hui2017}. Dark matter models with ultralight particles receive different names in the literature with popular ones being axion-like particles, wave dark matter or fuzzy dark matter. Here we refer to this model as $\psi$DM. In this type of models, the smooth macromodel gets perturbed by parsec-scale fluctuations (in galaxy clusters, and larger scales in galaxies) from the $\psi$DM field that result in big changes in the magnification near the CC \citep{Amruth2023,Diego2024c,Broadhurst2025,Palencia2025a}. These fluctuations create subparsec-scale caustic regions in the entire source plane, thus allowing again for the possibility of Hedorah being a very small group of stars with its counterimage being demagnified by a $\psi$DM fluctuation. This is graphically shown in the inset of Fig.~\ref{Fig_PDF_WaveDM} where the yellow circle represents a very small $r=0.1$ pc} star cluster in the source plane with the same macromodel magnification as the one expected for Hedorah's counterimage ($\mu_{\rm macro}=360$). For this particular case we use a model with de Broglie wavelength $\lambda_{\rm dB}=10$ pc, which is the scale expected for a massive galaxy cluster such as AS1063 with an axion mass $m_{\psi}=10^{-22}$ eV, and followed \cite{Diego2024c} to produce the simulated caustics. The projected distribution of $\psi$DM perturbations along the line of sight to Hedorah results in parsec scale $\psi$DM fluctuations, with dispersion in the projected mass $\sigma_{\psi}=4\, \Msun$ pc$^{-2}$  or $\approx 0.3\%$ of the macromodel value. This is consistent with the expected level of perturbation for the adopted value of $\lambda_{\rm dB}=10$ pc based on the central limit theorem and a cluster scale halo, where the smaller the value of $\lambda_{\rm dB}$, the more fluctuations are added along the line of sight, and the smaller the final fluctuations in the convergence.  
At this level of fluctuation, the value $\mu_{\rm macro}\sigma_{\psi}$ is comparable to the value of $\Sigma_{\rm crit}$ so we do expect the formation of critical curves around the $\psi$DM fluctuations. The level of distortion in the magnification is shown in  Fig.~\ref{Fig_PDF_WaveDM}, where we observe  a broad peak in the probability of magnification around the macromodel value. Interestingly, there is a tail of high magnification extending beyond $\mu=1000$ and an important increase in the probability of demagnification for images formed on either side of the critical curve (positive parity in blue, and negative parity in red). This symmetric behavior in counterimages with different parity is a unique feature of $\psi$DM models \cite{Diego2024c}. The large spread in magnification expected in this model helps explain the lack of counterimage for Hedorah in the scenario where Hedorah is a very compact group of stars, but the scale of the millicaustics is very small for this particular model making it challenging for structures with radii larger than 0.1 pc to be effectively demagnified. A larger scale of the millicaustics can be achieved if the de Broglie wavelength is larger than the one considered here (or similarly the axion mass is smaller), although lower masses for the axion are difficult to reconcile with some observations. 
If we allow for smaller axion masses, the $\psi$DM model reduces the tension in the CDM scenario regarding the low probability of having a millilens exactly aligned with Hedorah's counterimage. However the issue of its peculiar color still remains, making it harder to reconcile with a group of stars of different temperatures. 
Regarding $\psi$DM and Hedorah, this model is also helpful to magnify Hedorah more than in the CDM case. Combined with a microlens, the magnification for Hedorah could surpass the factor $\mu=5000$ required for a long period Cepheid to be observed at apparent magnitude AB$\approx 29.5$ in F356W. In $\psi$DM, these magnifications can be sustained during a microlensing event for longer periods, but still with a low overall probability, so we find that in general $\psi$DM does not provide conditions favorable enough to allow Hedorah to be interpreted as a classic long period Cepheid and a brighter yellow supergiant or yellow hypergiant is a more likely explanation for Hedorah. But even in this case, Hedorah still requires a microlens to temporarily boost its magnification, so if this interpretation is correct, future deep observations of AS1063 will probably not show Hedorah.

\section{Conclusions}\label{sect_conclusions}
We derived a new lens model for AS1063 from previously confirmed spectroscopic systems. Aided by the new lens model, we identified missing counterimages for some of the spectroscopically confirmed systems and identified entirely new system candidates in the very deep JWST images from the GLIMPSE program. We also confirm some of the previous system candidates via high consistency with the lens model, and color plus morphology match in the new JWST images. Positions, geometric redshifts, time delays, and magnifications are provided for all counterimages. 

We investigate in more detail a subsample of galaxies having very high magnification or at high redshift. One of these galaxies is very elongated, suggesting it is strongly lensed, but lacking obvious counterimages. We concluded that this galaxy is singly lensed but located at redshift $z=1.85$ in a region where magnification from the cluster plus a small subgroup of galaxies is very high ($\mu>100$) but without a caustic crossing the galaxy. The lensed image of the galaxy provides a full view of the entire galaxy (without a caustic slicing through the galaxy) but with differential magnification ranging from a few tens at the edges to over 200 at the center. We also discuss a high-redshift galaxy candidate at $z\approx 7.5$ that is triply lensed and showing intense H$\beta$ and/or OIII emission. The galaxy is slightly resolved in one of the counterimages allowing to estimate its size to $R\approx 150$ pc. It resembles an LRD with intense OIII emission and correspondingly modest Balmer break, and is also consistent with a small galaxy at $z\approx 7.5$ experiencing intense star formation. The multiply lensed nature of this galaxy and likely high star formation rate makes it a prime candidate for future follow up, in particular to monitor for possible supernovae.

We search for candidates to lensed stars and identified several candidates in a previously known $z=0.73$ caustic crossing galaxy that already showed microlensing events in past HST observations. The new candidates are found near the cluster CC, as expected for this type of objects, but we can not confirm their nature due to lack of a second epoch of similar depth. Also, proximity to the cluster center makes identification of microlensing candidates particularly challenging given the high density of unresolved GCs found at these distances from the center. Similar candidates are found in other low redshift caustic crossing galaxies ($z\lesssim 1$). A second deep observation of AS1063 with a subset of the filters used in GLIMPSE would confirm some of the candidates as microlensing events since these events are relatively short lived (weeks to months) and a second epoch should show some sources that disappear in the new  epoch as well as new sources appearing. Despite the lack of confirmation via the expected variability, the colors of some of the objects found near the critical curve are consistent with the expected colors of red supergiants at redshift $z=0.73$, and their apparent magnitudes are consistent with red supergiant stars being magnified during microlensing events. We estimated that stars from the TRGB should be detectable during microlensing events in the GLIMPSE data, opening the possibility to attempt an alternative distance estimation to the host galaxy based on the TRGB or J-AGB methods.

One of the caustic crossing galaxies at $z=3.72$ and farther away from the cluster center (i.e with far fewer GCs nearby) shows a point source, nicknamed Hedorah, with unique colors. The source is observed in one of the images of the galaxy but not in the other.
We explore several possibilities for Hedorah. Due to its very peculiar color, Hedorah is very unlikely a GC in the lens plane or behind the lens up to $z=1$. A small galaxy above $z=1$ would have a counterimage with similar unique colors in predictable positions which is not found, thus ruling out this possibility. For similar reasons, Hedorah is unlikely a small star forming region or globular cluster in the strongly lensed arc at $z=3.72$ hosting Hedorah since no counterimages are found. A small parsec-scale region in the host galaxy of Hedorah could be effectively hidden in one of the counterimages by a millilens, but we estimated this possibility to be very small, although not impossible. However, even at this redshift it is difficult to reconcile Hedorah's color with a group of stars, in particular if the group is older than a few tens of Myr, since in this case the 1.6 $\mu$m bump would be redshifted at higher wavelengths than the ones observed by GLIMPSE data, and Hedorah should show a much redder color than the one observed. Among the possibilities considered, we conclude that Hedorah is most likely a yellow monster, or kaiju,  star at $z=3.72$ (yellow supergiant star or yellow hypergiant star). Even these stars are often not bright enough to be detected at 0.18" from the CC by JWST, so Hedorah must have experienced a microlensing episode with increased luminosity during the JWST observations. Microlensing of stars near the cluster CC should be relatively frequent and in this scenario we expect to see only one of the counterimages, in agreement with the observations. 

Since many Cepheid stars are yellow supergiants, Hedorah could be the first example of a Cepheid star at cosmological ($z>0.1$) distances. We considered a bright Cepheid with period 100 days and find that even for this case, it would be challenging for a Cepheid at $z=3.72$ to be detectable during a microlensing event, but not impossible. A yellow hypergiant star is brighter and easier to detect at this redshift, and hence more likely. Despite the fact that the Cepheid interpretation is not the most likely, we can not rule out that possibility since the magnification can be higher than in the simple scenario considered in this work. To confirm the Cepheid nature of Hedorah, one needs to first establish its periodic nature and measure its period. This requires deep exposures similar to those of the GLIMPSE program. Confirming Hedorah is a Cepheid star at $z\approx3.7$ would be an incredible discovery, since Cepheid stars are of paramount importance for cosmology, occupying one of the first steps in the cosmic distance ladder. If Hedorah is not a Cepheid star and instead it is a yellow hypergiant, it could be approaching its final stages before going supernova (as for instance IRC+10420), thus offering the unique opportunity to study a SN precursor at $z>3$. Even in the case where Hedorah is not a single star at all, but a small group of stars, it is an interesting object to study in more detail. The fact that we do not observe its counterimage is intriguing and could be a consequence of dark matter being more exotic than the standard cold dark matter model. We have shown how fuzzy dark matter makes it more likely to observe one counterimage but not the other. Confirming one scenario (single monster star) vs the other (group of stars) is relatively simple because if Hedorah is a single star it requires the magnification boost provided by microlensing, that should show rapid changes in flux (weeks to months) while in the stellar group case, no changes (or percent level changes) are expected over time. A second deep observation with JWST reaching AB 30 mag in F200W and F277W should suffice to bring more clarity to this intriguing object. 

\section{Data availability}
The lens model (deflection field, convergence, shear and potential) is available at the link below. 
The model is computed for $D_{\rm ds}/D_{\rm s}=1$. Magnification maps at different redshifts are also available at the same link. 

https://zenodo.org/records/17726321


\begin{acknowledgements}
In honor of Mitch Struble, an always curious and gentle soul.
J.M.D. acknowledges the support of projects PID2022-138896NB-C51 (MCIU/AEI/MINECO/FEDER, UE) Ministerio de Ciencia, Investigaci\'on y Universidades and SA101P24. 

This work is based on observations made with the NASA/ESA/CSA \textit{James Webb} Space Telescope. The data were obtained from the Mikulski Archive for Space Telescopes at the Space Telescope Science Institute, which is operated by the Association of Universities for Research in Astronomy, Inc., under NASA contract NAS 5-03127 for JWST. These observations are associated with JWST programs \#3293.  
All the JWST data used in this paper can be found in MAST: {https://www.doi.org/10.17909/01wz-ng35}.

\end{acknowledgements}

\bibliographystyle{aa} 
\bibliography{MyBiblio} 

\begin{appendix}

\section{\label{sc_appendixA}Lensing constraints}
The Rank A systems listed below were compiled as part of the BUFFALO lens modeling effort \citep{Steinhardt2020}. The data set is a compilation of systems from the literature \citep{Caminha2016,Diego2016b,Karman2017,Kawamata2018}. Details of the grading and classification system followed by the BUFFALO team can be found in \cite{Niemiec2023}. We also include three systems (35, 36, and 37 in Table~\ref{tab_arcs}) from \cite{Beauchesne2024}, for which we identify two additional missing counterimages. Many of the B rank systems were also compiled by the BUFFALO collaboration. New systems or redefinitions based on the new JWST and new lens model are marked with superindices next to the arc Id. New systems identified in the JWST images (and validated by the lens model) are also included. Only systems with Rank A were used to derive the lens model. Only the most reliable rank B systems are included here. Rank B systems that appear surrounded by Rank A constraints should have reliable geometric redshifts \zg. The redshifts of the rank B systems that are farther away from the center of the cluster and outside the region constrained by the Rank A systems may be less reliable due to the biased-low masses in the lens model in the outer regions. The lower mass in the outer regions is expected to bias the \zg\, high. This bias explains the apparently high \zg\, estimates for the LRD candidate and the triply lensed $z\sim10$ galaxy (systems 78 and 80 respectively). Magnification factors and time delays are computed at the observed positions. Time delays (in years) are presented with respect to the image arriving first. Uncertainties (1$\sigma$) are derived from the same range of models (blue band) shown in Figure~\ref{Fig_Profiles}.

%
\onecolumn
\begin{longtable}{rcccccrc}
\caption{Lensed families and multiple images used to derive the lens model (total of 263 counterimages, including knots, from 61 individual galaxies and 3 giant arcs) RA and DEC are given in the J2000 epoch. 
\label{tab_arcs}}\\
\hline\hline
  \textbf{ID}
& \textbf{RA}
& \textbf{DEC}
& \textbf{\zs}
& \textbf{\zg}
& \textbf{$\mu$}
& \textbf{$\delta T$ (yr)}
& \textbf{Rank} \\
\hline
\endfirsthead
\caption{Continued on next page.}\\
\hline\hline
  \textbf{ID}
& \textbf{RA}
& \textbf{DEC}
& \textbf{\zs}
& \textbf{\zg}
& \textbf{$\mu$}
& \textbf{$\delta T$ (yr)}
& \textbf{Rank} \\
\hline
\endhead
\hline
\endfoot
\hline
\endlastfoot
\hline
      1.a & 342.1944580 & -44.5270042 &  1.2290 &  1.22$\pm$0.03 &   9.2$\pm$  1.0 &   6.54$\pm$  0.35 & A  \\
      1.b & 342.1958618 & -44.5289497 &    -    &         -        &  23.6$\pm$  9.5 &   6.05$\pm$  0.43 & A  \\
      1.c & 342.1864319 & -44.5212021 &    -    &         -        &   5.3$\pm$  0.4 &   0.00$\pm$  0.00 & A  \\
\hline
      2.a & 342.1948242 & -44.5273552 &  1.2290 &  1.23$\pm$0.03 &  22.1$\pm$  3.9 &   7.36$\pm$  0.40 & A  \\
      2.b & 342.1955872 & -44.5284042 &    -    &         -        &  52.1$\pm$ 38.2 &   7.15$\pm$  0.49 & A  \\
      2.c & 342.1864014 & -44.5211296 &    -    &         -        &   5.1$\pm$  0.4 &   0.00$\pm$  0.00 & A  \\
\hline
      3.a & 342.1952209 & -44.5278244 &  1.2290 &  1.24$\pm$0.03 & 134.1$\pm$ 37.5 &   7.47$\pm$  0.53 & A  \\
      3.b & 342.1952820 & -44.5279121 &    -    &         -        &  95.7$\pm$  3.2 &   7.48$\pm$  0.53 & A  \\
      3.c & 342.1863098 & -44.5210533 &    -    &         -        &   5.0$\pm$  0.4 &   0.00$\pm$  0.00 & A  \\
\hline
      4.a & 342.1940613 & -44.5265770 &  1.2290 &  1.22$\pm$0.03 &   3.9$\pm$  0.5 &   5.12$\pm$  0.37 & A  \\
      4.b & 342.1962585 & -44.5296631 &    -    &         -        &  12.2$\pm$  2.8 &   4.16$\pm$  0.41 & A  \\
      4.c & 342.1867676 & -44.5213547 &    -    &         -        &   5.8$\pm$  0.5 &   0.00$\pm$  0.00 & A  \\
\hline
      5.a & 342.1927185 & -44.5311890 &  1.2610 &  1.27$\pm$0.04 &  25.9$\pm$  4.7 &  24.28$\pm$  0.83 & A  \\
      5.b & 342.1921387 & -44.5298309 &    -    &         -        &  22.4$\pm$  0.4 &  24.69$\pm$  0.93 & A  \\
      5.c & 342.1798706 & -44.5215607 &    -    &         -        &   3.6$\pm$  0.5 &   0.00$\pm$  0.00 & A  \\
\hline
      6.a & 342.1925659 & -44.5307350 &  1.2610 &  1.27$\pm$0.03 &  91.2$\pm$ 37.1 &  24.91$\pm$  0.79 & A  \\
      6.b & 342.1923828 & -44.5302696 &    -    &         -        &  66.8$\pm$  4.7 &  24.97$\pm$  0.87 & A  \\
      6.c & 342.1797485 & -44.5215759 &    -    &         -        &   3.6$\pm$  0.5 &   0.00$\pm$  0.00 & A  \\
\hline
      7.a & 342.1924438 & -44.5309067 &  1.2610 &  1.26$\pm$0.03 &  58.2$\pm$ 14.6 &  25.29$\pm$  0.91 & A  \\
      7.b & 342.1921997 & -44.5303154 &    -    &         -        &  58.8$\pm$  3.5 &  25.47$\pm$  0.92 & A  \\
      7.c & 342.1796875 & -44.5216293 &    -    &         -        &   3.6$\pm$  0.5 &   0.00$\pm$  0.00 & A  \\
\hline
      8.a & 342.1936951 & -44.5301628 &  1.2583 &  1.25$\pm$0.03 &  90.0$\pm$ 63.5 &  17.95$\pm$  0.22 & A  \\
      8.b & 342.1933289 & -44.5294189 &    -    &         -        &  26.4$\pm$  2.8 &  17.94$\pm$  0.24 & A  \\
   8.c\ma & 342.1816406 & -44.5214043 &    -    &         -        &   3.9$\pm$  0.5 &   0.00$\pm$  0.00 & A  \\
\hline
      9.a & 342.1791992 & -44.5235901 &  1.3980 &  1.41$\pm$0.04 &   5.0$\pm$  0.8 &  11.20$\pm$  0.08 & A  \\
      9.b & 342.1878357 & -44.5273094 &    -    &         -        &   4.5$\pm$  0.4 &  21.00$\pm$  0.08 & A  \\
      9.c & 342.1931763 & -44.5365295 &    -    &         -        &   3.8$\pm$  0.2 &   0.00$\pm$  0.00 & A  \\
\hline
     10.a & 342.1788940 & -44.5236092 &  1.3980 &  1.41$\pm$0.04 &   4.9$\pm$  0.8 &   9.55$\pm$  0.07 & A  \\
     10.b & 342.1877136 & -44.5275040 &    -    &         -        &   4.3$\pm$  0.4 &  20.49$\pm$  0.09 & A  \\
     10.c & 342.1929016 & -44.5365944 &    -    &         -        &   3.9$\pm$  0.2 &   0.00$\pm$  0.00 & A  \\
\hline
     11.a & 342.1782227 & -44.5236626 &  1.3980 &  1.41$\pm$0.04 &   5.1$\pm$  1.1 &   6.36$\pm$  0.09 & A  \\
     11.b & 342.1875305 & -44.5278740 &    -    &         -        &   5.4$\pm$  0.6 &  19.35$\pm$  0.08 & A  \\
     11.c & 342.1923828 & -44.5367279 &    -    &         -        &   4.0$\pm$  0.3 &   0.00$\pm$  0.00 & A  \\
\hline
     12.a & 342.1742554 & -44.5283318 &  1.4280 &  1.44$\pm$0.05 &   6.6$\pm$  0.6 &  29.98$\pm$  0.49 & A  \\
     12.b & 342.1758423 & -44.5325394 &    -    &         -        &   9.2$\pm$  1.7 &  31.62$\pm$  0.24 & A  \\
     12.c & 342.1884460 & -44.5399933 &    -    &         -        &   4.2$\pm$  0.5 &   0.00$\pm$  0.00 & A  \\
\hline
     13.a & 342.1741943 & -44.5286293 &  1.4280 &  1.44$\pm$0.05 &   7.4$\pm$  0.8 &  31.76$\pm$  0.50 & A  \\
     13.b & 342.1756287 & -44.5323296 &    -    &         -        &   9.5$\pm$  1.5 &  32.89$\pm$  0.25 & A  \\
     13.c & 342.1883850 & -44.5400963 &    -    &         -        &   4.6$\pm$  0.8 &   0.00$\pm$  0.00 & A  \\
\hline
     14.a & 342.1740417 & -44.5289001 &  1.4280 &  1.43$\pm$0.05 &  12.3$\pm$  2.9 &  31.93$\pm$  0.44 & A  \\
     14.b & 342.1753845 & -44.5322609 &    -    &         -        &   9.8$\pm$  1.4 &  32.83$\pm$  0.28 & A  \\
     14.c & 342.1881714 & -44.5402260 &    -    &         -        &   3.4$\pm$  0.1 &   0.00$\pm$  0.00 & A  \\
\hline
     15.a & 342.1740723 & -44.5276871 &  1.4280 &  1.45$\pm$0.05 &   5.4$\pm$  0.5 &  24.23$\pm$  0.53 & A  \\
     15.b & 342.1763306 & -44.5330620 &    -    &         -        &  12.9$\pm$  3.2 &  27.95$\pm$  0.32 & A  \\
     15.c & 342.1883545 & -44.5397873 &    -    &         -        &   3.8$\pm$  0.2 &   0.00$\pm$  0.00 & A  \\
\hline
     16.a & 342.1740417 & -44.5274696 &  1.4280 &  1.45$\pm$0.05 &   5.1$\pm$  0.4 &  21.35$\pm$  0.48 & A  \\
     16.b & 342.1765442 & -44.5331497 &    -    &         -        &   4.8$\pm$  0.2 &  26.38$\pm$  0.33 & A  \\
     16.c & 342.1883240 & -44.5396767 &    -    &         -        &   3.7$\pm$  0.1 &   0.00$\pm$  0.00 & A  \\
\hline
     17.a & 342.1744385 & -44.5277710 &  1.4280 &  1.45$\pm$0.05 &   5.8$\pm$  0.5 &  27.84$\pm$  0.44 & A  \\
     17.b & 342.1763611 & -44.5326691 &    -    &         -        &   7.5$\pm$  1.2 &  30.81$\pm$  0.25 & A  \\
     17.c & 342.1889038 & -44.5397720 &    -    &         -        &   3.7$\pm$  0.2 &   0.00$\pm$  0.00 & A  \\
\hline
     18.a & 342.1693726 & -44.5272484 &  1.8370 &  1.84$\pm$0.10 &   3.5$\pm$  0.0 &   0.00$\pm$  0.00 & A  \\
     18.b & 342.1742554 & -44.5371094 &    -    &         -        &   6.0$\pm$  0.2 &  13.91$\pm$  0.12 & A  \\
     18.c & 342.1818237 & -44.5405006 &    -    &         -        &  30.3$\pm$ 15.0 &  10.02$\pm$  0.56 & A  \\
\hline
     19.a & 342.1902466 & -44.5297623 &  0.7300 &  0.74$\pm$0.03 &  27.8$\pm$  4.0 &   0.70$\pm$  0.21 & A  \\
     19.b & 342.1895447 & -44.5288429 &    -    &         -        &  71.0$\pm$  4.2 &   0.60$\pm$  0.10 & A  \\
     19.c & 342.1849670 & -44.5257034 &    -    &         -        &   8.5$\pm$  0.1 &   0.00$\pm$  0.00 & A  \\
\hline
     20.a & 342.1902161 & -44.5303574 &  0.7300 &  0.74$\pm$0.03 &  10.5$\pm$  0.1 &   0.26$\pm$  0.16 & A  \\
     20.b & 342.1888428 & -44.5285225 &    -    &         -        & 163.1$\pm$ 63.8 &   1.08$\pm$  0.23 & A  \\
  20.c\ma & 342.1852722 & -44.5256157 &    -    &         -        &   8.4$\pm$  0.1 &   0.05$\pm$  0.08 & A  \\
\hline
     21.a & 342.1896973 & -44.5297546 &  0.7300 &  0.72$\pm$0.03 & 154.9$\pm$ 76.9 &   2.19$\pm$  0.05 & A  \\
     21.b & 342.1896362 & -44.5295753 &    -    &         -        &   3.9$\pm$  0.8 &   2.19$\pm$  0.12 & A  \\
     21.c & 342.1847534 & -44.5256767 &    -    &         -        &   8.0$\pm$  0.0 &   0.00$\pm$  0.00 & A  \\
\hline
     22.a & 342.1750488 & -44.5410309 &  3.1160 &  3.00$\pm$0.41 &  31.2$\pm$  3.8 &   9.14$\pm$  0.26 & A  \\
     22.b & 342.1731567 & -44.5399818 &    -    &         -        &  10.3$\pm$  1.9 &   9.50$\pm$  0.31 & A  \\
     22.c & 342.1655579 & -44.5295296 &    -    &         -        &   4.8$\pm$  0.4 &   0.00$\pm$  0.00 & A  \\
\hline
     23.a & 342.1890564 & -44.5300484 &  6.1120 &  6.38$\pm$2.16 &  52.2$\pm$ 63.4 &  78.52$\pm$  0.39 & A  \\
     23.b & 342.1810303 & -44.5346184 &    -    &         -        &   9.4$\pm$  3.0 &  77.95$\pm$  0.41 & A  \\
     23.c & 342.1908875 & -44.5374680 &    -    &         -        &   7.7$\pm$  1.3 &  70.36$\pm$  0.20 & A  \\
     23.d & 342.1712952 & -44.5198097 &    -    &         -        &   2.3$\pm$  0.1 &   0.00$\pm$  0.00 & A  \\
     23.e & 342.1840820 & -44.5316277 &    -    &         -        &   2.2$\pm$  0.2 &  79.08$\pm$  0.46 & A  \\
\hline
     24.a & 342.1815491 & -44.5393753 &  4.1130 &  4.14$\pm$0.05 &  24.4$\pm$  1.0 &  60.07$\pm$  1.29 & A  \\
     24.b & 342.1791382 & -44.5386772 &    -    &         -        &  26.5$\pm$  0.6 &  59.84$\pm$  1.21 & A  \\
  24.c\ma & 342.1687927 & -44.5223770 &    -    &         -        &   2.3$\pm$  0.0 &   0.00$\pm$  0.00 & A  \\
\hline
     25.a & 342.1788330 & -44.5358696 &  3.1180 &  3.10$\pm$0.28 &   4.9$\pm$  0.2 &  54.78$\pm$  0.07 & A  \\
     25.b & 342.1874084 & -44.5386887 &    -    &         -        &   7.3$\pm$  0.8 &  49.81$\pm$  0.31 & A  \\
     25.c & 342.1706543 & -44.5220871 &    -    &         -        &   2.6$\pm$  0.1 &   0.00$\pm$  0.00 & A  \\
\hline
     26.a & 342.1796265 & -44.5357208 &  3.1180 &  3.13$\pm$0.30 &   6.6$\pm$  0.6 &  59.82$\pm$  0.08 & A  \\
     26.b & 342.1877747 & -44.5382805 &    -    &         -        &   7.8$\pm$  1.1 &  55.81$\pm$  0.31 & A  \\
     26.c & 342.1708374 & -44.5216904 &    -    &         -        &   2.5$\pm$  0.1 &   0.00$\pm$  0.00 & A  \\
\hline
     27.a & 342.1799927 & -44.5356026 &  3.1180 &  3.12$\pm$0.26 &   7.8$\pm$  1.2 &  62.46$\pm$  0.08 & A  \\
     27.b & 342.1878662 & -44.5380402 &    -    &         -        &   8.2$\pm$  1.3 &  58.74$\pm$  0.29 & A  \\
     27.c & 342.1709290 & -44.5214958 &    -    &         -        &   2.5$\pm$  0.1 &   0.00$\pm$  0.00 & A  \\
\hline
     28.a & 342.1782532 & -44.5360031 &  3.1180 &  3.10$\pm$0.29 &   3.5$\pm$  0.1 &  50.23$\pm$  0.04 & A  \\
     28.b & 342.1871338 & -44.5390434 &    -    &         -        &   6.7$\pm$  0.5 &  44.60$\pm$  0.21 & A  \\
     28.c & 342.1704712 & -44.5224648 &    -    &         -        &   2.6$\pm$  0.1 &   0.00$\pm$  0.00 & A  \\
\hline
     29.a & 342.1782532 & -44.5358238 &  3.1180 &  3.15$\pm$0.28 &   4.0$\pm$  0.0 &  49.73$\pm$  0.07 & A  \\
     29.b & 342.1876831 & -44.5390015 &    -    &         -        &   6.5$\pm$  0.6 &  43.01$\pm$  0.11 & A  \\
     29.c & 342.1705933 & -44.5223846 &    -    &         -        &   2.7$\pm$  0.1 &   0.00$\pm$  0.00 & A  \\
\hline
     30.a & 342.1858215 & -44.5388489 &  3.6060 &  3.53$\pm$0.40 &   9.5$\pm$  1.0 &  59.44$\pm$  0.91 & A  \\
     30.b & 342.1788330 & -44.5367317 &    -    &         -        &   5.7$\pm$  0.1 &  61.32$\pm$  0.61 & A  \\
  30.c\ma & 342.1697998 & -44.5219765 &    -    &         -        &   2.4$\pm$  0.0 &   0.00$\pm$  0.00 & A  \\
\hline
     31.a & 342.1800232 & -44.5384293 &  1.0350 &  1.02$\pm$0.03 &   6.8$\pm$  0.1 &   1.60$\pm$  0.24 & A  \\
     31.b & 342.1755371 & -44.5359383 &    -    &         -        &  48.3$\pm$ 36.4 &   3.35$\pm$  0.31 & A  \\
     31.c & 342.1719055 & -44.5302505 &    -    &         -        &   5.8$\pm$  0.6 &   0.00$\pm$  0.00 & A  \\
\hline
     32.a & 342.1809082 & -44.5387344 &  1.0350 &  1.02$\pm$0.03 &   5.5$\pm$  0.1 &   0.00$\pm$  0.00 & A  \\
     32.b & 342.1722717 & -44.5307045 &    -    &         -        &   7.5$\pm$  0.9 &   2.16$\pm$  0.07 & A  \\
\hline
     33.a & 342.1803284 & -44.5387154 &  1.0350 &  1.03$\pm$0.03 &   5.9$\pm$  0.1 &   0.00$\pm$  0.00 & A  \\
     33.b & 342.1719055 & -44.5307159 &    -    &         -        &   6.8$\pm$  0.8 &   0.53$\pm$  0.08 & A  \\
\hline
     34.a & 342.1788940 & -44.5380211 &  1.0350 &  1.02$\pm$0.03 &   8.8$\pm$  0.1 &   5.75$\pm$  0.11 & A  \\
     34.b & 342.1715698 & -44.5299568 &    -    &         -        &   5.0$\pm$  0.5 &   0.00$\pm$  0.00 & A  \\
\hline
     35.a & 342.1758118 & -44.5363693 &  3.6660 &  3.48$\pm$0.31 &   9.9$\pm$  3.0 &  32.97$\pm$  0.65 & A  \\
     35.b & 342.1878967 & -44.5404205 &    -    &         -        &   3.9$\pm$  0.2 &  18.11$\pm$  1.31 & A  \\
  35.c\ma & 342.1705322 & -44.5235825 &    -    &         -        &   3.1$\pm$  0.1 &   0.00$\pm$  0.00 & A  \\
\hline
     36.a & 342.1737976 & -44.5411797 &  3.2300 &  3.20$\pm$0.37 &  26.6$\pm$  2.8 &   5.89$\pm$  0.55 & A  \\
     36.b & 342.1739197 & -44.5412369 &    -    &         -        &  25.4$\pm$  4.2 &   5.76$\pm$  0.55 & A  \\
  36.c\ma & 342.1650696 & -44.5310555 &    -    &         -        &   6.8$\pm$  1.5 &   0.00$\pm$  0.00 & A  \\
\hline
     37.a & 342.1802979 & -44.5246391 &  3.0820 &  3.38$\pm$0.05 &  15.0$\pm$  4.2 &   0.00$\pm$  0.00 & A  \\
     37.b & 342.1838684 & -44.5257568 &    -    &         -        &  14.4$\pm$  2.9 &   0.71$\pm$  0.10 & A  \\
\hline
     38.a & 342.1983643 & -44.5357513 &  2.9760 &  2.98$\pm$0.27 &   3.0$\pm$  0.1 &   0.00$\pm$  0.00 & A  \\
     38.b & 342.1924438 & -44.5250702 &    -    &         -        &   5.3$\pm$  0.6 &  22.41$\pm$  1.22 & A  \\
     38.c & 342.1815186 & -44.5202599 &    -    &         -        &  13.4$\pm$  5.4 &  12.51$\pm$  1.94 & A  \\
\hline
  39.a\mc & 342.1828308 & -44.5202789 &  3.1690 &  3.30$\pm$0.13 &   5.1$\pm$  0.7 &   0.00$\pm$  0.00 & A  \\
  39.b\mc & 342.1919556 & -44.5240898 &    -    &         -        &   5.7$\pm$  0.7 &   5.45$\pm$  0.21 & A  \\
\hline
     40.a & 342.1901550 & -44.5309296 &  5.0510 &  3.97$\pm$1.61 &  13.8$\pm$  4.6 &   1.25$\pm$  0.39 & A  \\
     40.b & 342.1908569 & -44.5356598 &    -    &         -        &  12.7$\pm$  3.2 &   0.00$\pm$  0.00 & A  \\
\hline
     41.a & 342.1837769 & -44.5212479 &  5.2373 &  4.56$\pm$2.40 &  11.6$\pm$  3.4 &   0.00$\pm$  0.00 & A  \\
     41.b & 342.1887512 & -44.5227585 &    -    &         -        &  11.1$\pm$  2.2 &   1.34$\pm$  0.13 & A  \\
\hline
  42.a\mc & 342.1970215 & -44.5221214 &  5.8940 &  5.02$\pm$1.86 &   9.0$\pm$  0.3 &   1.98$\pm$  0.17 & A  \\
  42.b\mc & 342.1900940 & -44.5178871 &    -    &         -        &   7.0$\pm$  0.9 &   0.00$\pm$  0.00 & A  \\
\hline
     43.a & 342.1960754 & -44.5230103 &  3.2857 &  3.26$\pm$0.38 &  10.2$\pm$  0.5 &  11.67$\pm$  1.50 & A  \\
     43.b & 342.1891785 & -44.5187225 &    -    &         -        &   6.9$\pm$  0.9 &   9.23$\pm$  1.60 & A  \\
  43.c\ma & 342.2021484 & -44.5321312 &    -    &         -        &   3.3$\pm$  0.2 &   0.00$\pm$  0.00 & A  \\
\hline
     44.a & 342.1955566 & -44.5321388 & -1.0000 &  2.07$\pm$0.08 &  11.8$\pm$  1.9 &  34.90$\pm$  0.08 & B  \\
     44.b & 342.1939087 & -44.5287323 &    -    &         -        &   7.7$\pm$  0.1 &  35.70$\pm$  0.08 & B  \\
     44.c & 342.1785583 & -44.5195465 &    -    &         -        &   3.2$\pm$  0.3 &   0.00$\pm$  0.00 & B  \\
\hline
     45.a & 342.1679688 & -44.5261993 & -1.0000 &  2.61$\pm$0.22 &   3.2$\pm$  0.1 &   0.00$\pm$  0.00 & B  \\
     45.b & 342.1746216 & -44.5383873 &    -    &         -        &   5.2$\pm$  0.3 &  22.35$\pm$  0.14 & B  \\
     45.c & 342.1807861 & -44.5408821 &    -    &         -        &   6.6$\pm$  0.4 &  20.68$\pm$  0.08 & B  \\
\hline
     46.a & 342.1677856 & -44.5262794 & -1.0000 &  2.62$\pm$0.22 &   3.0$\pm$  0.1 &   0.00$\pm$  0.00 & B  \\
     46.b & 342.1748047 & -44.5386124 &    -    &         -        &   5.6$\pm$  0.5 &  22.97$\pm$  0.34 & B  \\
     46.c & 342.1802979 & -44.5408096 &    -    &         -        &   7.3$\pm$  0.3 &  21.32$\pm$  0.04 & B  \\
\hline
     47.a & 342.1917114 & -44.5305176 & -1.0000 &  3.15$\pm$0.27 &   9.9$\pm$  1.5 &  70.17$\pm$  0.36 & B  \\
     47.b & 342.1925354 & -44.5344200 &    -    &         -        &  13.4$\pm$  3.1 &  69.30$\pm$  0.07 & B  \\
     47.c & 342.1736755 & -44.5194016 &    -    &         -        &   2.5$\pm$  0.1 &   0.00$\pm$  0.00 & B  \\
\hline
     48.a & 342.1930542 & -44.5355301 & -1.0000 &  2.78$\pm$0.20 &   6.9$\pm$  0.8 &  48.76$\pm$  0.52 & B  \\
     48.b & 342.1912537 & -44.5292206 &    -    &         -        &   5.0$\pm$  0.5 &  53.57$\pm$  0.65 & B  \\
     48.c & 342.1744690 & -44.5200844 &    -    &         -        &   2.9$\pm$  0.2 &   0.00$\pm$  0.00 & B  \\
\hline
     49.a & 342.1974487 & -44.5306320 & -1.0000 &  2.11$\pm$0.11 &  13.3$\pm$  2.1 &  25.44$\pm$  0.86 & B  \\
     49.b & 342.1963806 & -44.5283928 &    -    &         -        &  25.4$\pm$  2.7 &  25.71$\pm$  1.03 & B  \\
  49.c\mc & 342.1816711 & -44.5187454 &    -    &         -        &   3.5$\pm$  0.3 &   0.00$\pm$  0.00 & B  \\
\hline
     50.a & 342.1919250 & -44.5368233 & -1.0000 &  1.26$\pm$0.03 &   3.8$\pm$  0.3 &   0.00$\pm$  0.00 & B  \\
     50.b & 342.1862488 & -44.5280685 &    -    &         -        &   3.3$\pm$  0.0 &  24.02$\pm$  0.41 & B  \\
  50.c\ma & 342.1790466 & -44.5246048 &    -    &         -        &  15.1$\pm$ 10.5 &  16.70$\pm$  0.33 & B  \\
  50.d\ma & 342.1787109 & -44.5249214 &    -    &         -        &   7.6$\pm$  2.3 &  16.69$\pm$  0.30 & B  \\
  50.e\mc & 342.1826477 & -44.5307045 &    -    &         -        &   1.8$\pm$  0.1 &  26.78$\pm$  1.03 & B  \\
\hline
     51.a & 342.1869812 & -44.5274162 & -1.0000 &  0.75$\pm$0.03 &  11.8$\pm$  0.1 &   2.54$\pm$  0.37 & B  \\
     51.b & 342.1852112 & -44.5263405 &    -    &         -        &  26.0$\pm$  1.5 &   2.17$\pm$  0.13 & B  \\
     51.c & 342.1865234 & -44.5271339 &    -    &         -        &  99.3$\pm$ 10.3 &   2.61$\pm$  0.38 & B  \\
     51.d & 342.1860352 & -44.5268402 &    -    &         -        & 121.8$\pm$ 47.2 &   2.49$\pm$  0.35 & B  \\
     51.e & 342.1856384 & -44.5265846 &    -    &         -        &  42.1$\pm$  7.4 &   2.30$\pm$  0.18 & B  \\
     51.f & 342.1903381 & -44.5315475 &    -    &         -        &   6.6$\pm$  0.1 &   0.00$\pm$  0.00 & B  \\
\hline
     52.a & 342.2036743 & -44.5199776 & -1.0000 &  1.69$\pm$1.45 &  18.9$\pm$ 15.6 &   0.00$\pm$  0.00 & B  \\
     52.b & 342.2032776 & -44.5195122 &    -    &         -        &   8.0$\pm$  2.8 &   0.08$\pm$  0.02 & B  \\
     52.c & 342.2027283 & -44.5191345 &    -    &         -        &  25.4$\pm$ 29.3 &   0.35$\pm$  0.07 & B  \\
\hline
     53.a & 342.1708984 & -44.5295982 & -1.0000 &  3.07$\pm$0.30 &  13.5$\pm$  3.3 &  44.60$\pm$  2.29 & B  \\
     53.b & 342.1715393 & -44.5323639 &    -    &         -        &   9.4$\pm$  0.2 &  44.77$\pm$  2.14 & B  \\
  53.c\ma & 342.1887512 & -44.5426559 &    -    &         -        &   2.4$\pm$  0.2 &   0.00$\pm$  0.00 & B  \\
\hline
     54.a & 342.2115479 & -44.5260353 & -1.0000 &  1.15$\pm$0.46 &   2.3$\pm$  0.3 &   0.13$\pm$  0.06 & B  \\
     54.b & 342.2111206 & -44.5256767 &    -    &         -        &   1.4$\pm$  0.4 &   0.19$\pm$  0.09 & B  \\
     54.c & 342.2111511 & -44.5252342 &    -    &         -        &   3.3$\pm$  0.4 &   0.00$\pm$  0.00 & B  \\
\hline
     55.a & 342.2028809 & -44.5252953 & -1.0000 &  2.07$\pm$0.45 &  40.0$\pm$ 16.3 &   0.00$\pm$  0.00 & B  \\
     55.b & 342.1996765 & -44.5215454 &    -    &         -        &  84.3$\pm$ 69.9 &   0.57$\pm$  0.53 & B  \\
     55.c & 342.1980896 & -44.5201187 &    -    &         -        &  45.8$\pm$ 45.8 &   0.38$\pm$  0.23 & B  \\
\hline
     56.a & 342.1753540 & -44.5397377 & -1.0000 &  1.63$\pm$0.08 &  38.5$\pm$  6.5 &   4.80$\pm$  0.12 & B  \\
     56.b & 342.1744690 & -44.5392380 &    -    &         -        &  25.4$\pm$  8.1 &   4.86$\pm$  0.07 & B  \\
     56.c & 342.1677856 & -44.5306778 &    -    &         -        &   6.0$\pm$  1.0 &   0.00$\pm$  0.00 & B  \\
\hline
     57.a & 342.1985474 & -44.5208054 & -1.0000 &  2.80$\pm$0.36 &  48.1$\pm$ 23.7 &   3.39$\pm$  1.41 & B  \\
     57.b & 342.1972961 & -44.5197678 &    -    &         -        &  69.5$\pm$ 75.3 &   3.30$\pm$  1.15 & B  \\
  57.c\ma & 342.2043152 & -44.5277557 &    -    &         -        &   4.8$\pm$  0.5 &   0.00$\pm$  0.00 & B  \\
\hline
     58.a & 342.1983032 & -44.5205841 & -1.0000 &  2.85$\pm$0.52 &  49.0$\pm$ 22.6 &   0.05$\pm$  0.09 & B  \\
     58.b & 342.1974792 & -44.5199280 &    -    &         -        &  50.2$\pm$  6.6 &   0.05$\pm$  0.05 & B  \\
\hline
     59.a & 342.1990051 & -44.5212288 & -1.0000 &  2.81$\pm$1.55 &  18.0$\pm$  5.0 &   0.15$\pm$  0.19 & B  \\
     59.b & 342.1967773 & -44.5194092 &    -    &         -        &  31.1$\pm$ 21.0 &   0.00$\pm$  0.01 & B  \\
\hline
     60.a & 342.1802368 & -44.5392647 & 3.72    &  3.08$\pm$0.21   &  30.6$\pm$  0.9 &  49.50$\pm$  1.09 & B  \\
     60.b & 342.1789246 & -44.5388603 &    -    &         -        &  49.5$\pm$  6.3 &  49.23$\pm$  0.96 & B  \\
  60.c\ma & 342.1684570 & -44.5235786 &    -    &         -        &   2.4$\pm$  0.1 &   0.00$\pm$  0.00 & B  \\
\hline
     61.a & 342.1910706 & -44.5233231 & -1.0000 &  1.74$\pm$0.03 &  15.0$\pm$  4.0 &  18.07$\pm$  0.57 & B  \\
     61.b & 342.1905212 & -44.5229568 &    -    &         -        &  27.7$\pm$ 13.6 &  18.04$\pm$  0.51 & B  \\
     61.c & 342.1895752 & -44.5224457 &    -    &         -        & 161.9$\pm$ 65.8 &  17.77$\pm$  0.43 & B  \\
     61.d & 342.1887512 & -44.5220680 &    -    &         -        &  25.5$\pm$  2.5 &  17.63$\pm$  0.33 & B  \\
  61.e\ma & 342.1994934 & -44.5333862 &    -    &         -        &   3.2$\pm$  0.1 &   0.00$\pm$  0.00 & B  \\
\hline
     62.a & 342.1547546 & -44.5395546 & -1.0000 &  0.83$\pm$0.09 &   1.3$\pm$  0.4 &   0.31$\pm$  0.10 & B  \\
     62.b & 342.1545715 & -44.5395355 &    -    &         -        &   0.6$\pm$  0.1 &   0.17$\pm$  0.10 & B  \\
     62.c & 342.1542969 & -44.5388680 &    -    &         -        &   1.4$\pm$  0.2 &   0.00$\pm$  0.00 & B  \\
     62.d & 342.1548767 & -44.5393906 &    -    &         -        &   0.2$\pm$  0.1 &   0.14$\pm$  0.10 & B  \\
\hline
     63.a & 342.1772766 & -44.5407028 & -1.0000 &  1.61$\pm$0.09 &   8.6$\pm$  1.7 &   0.92$\pm$  0.19 & B  \\
     63.b & 342.1716309 & -44.5371704 &    -    &         -        &  18.2$\pm$  5.6 &   2.18$\pm$  0.45 & B  \\
     63.c & 342.1680603 & -44.5314789 &    -    &         -        &   7.9$\pm$  1.8 &   0.00$\pm$  0.00 & B  \\
\hline
     64.a & 342.1988831 & -44.5347672 & -1.0000 &  2.64$\pm$0.19 &   3.2$\pm$  0.1 &   0.00$\pm$  0.00 & B  \\
     64.b & 342.1932983 & -44.5248413 &    -    &         -        &   6.5$\pm$  0.8 &  17.65$\pm$  1.36 & B  \\
     64.c & 342.1831055 & -44.5198898 &    -    &         -        &   4.5$\pm$  0.6 &   9.94$\pm$  1.68 & B  \\
\hline
     65.a & 342.1916809 & -44.5366898 & -1.0000 &  1.28$\pm$0.03 &   4.1$\pm$  0.3 &   0.00$\pm$  0.00 & B  \\
     65.b & 342.1865845 & -44.5282822 &    -    &         -        &   4.4$\pm$  0.3 &  20.41$\pm$  0.32 & B  \\
     65.c & 342.1784058 & -44.5244293 &    -    &         -        &   7.8$\pm$  3.3 &   9.70$\pm$  0.15 & B  \\
  65.d\ma & 342.1821899 & -44.5310974 &    -    &         -        &   2.2$\pm$  0.2 &  23.65$\pm$  0.89 & B  \\
\hline
     66.a & 342.1870728 & -44.5291100 & -1.0000 &  1.68$\pm$0.03 &   8.1$\pm$  2.0 &  26.39$\pm$  0.69 & B  \\
     66.b & 342.1810303 & -44.5324211 &    -    &         -        &   2.7$\pm$  0.0 &  27.43$\pm$  0.83 & B  \\
  66.c\ma & 342.1756592 & -44.5231972 &    -    &         -        &   4.1$\pm$  0.6 &   0.00$\pm$  0.00 & B  \\
     66.d & 342.1911316 & -44.5372887 &    -    &         -        &   4.7$\pm$  0.4 &   8.78$\pm$  0.71 & B  \\
\hline
  67.a\mb & 342.1834717 & -44.5264587 & -1.0000 &  3.06$\pm$0.68 &  33.9$\pm$ 14.0 &  83.15$\pm$  0.48 & B  \\
  67.b\mb & 342.1802979 & -44.5256958 &    -    &         -        &  26.4$\pm$  6.8 &  83.24$\pm$  0.90 & B  \\
  67.c\mc & 342.1959229 & -44.5405312 &    -    &         -        &   1.9$\pm$  0.1 &   0.00$\pm$  0.00 & B  \\
\hline
  68.a\mb & 342.1694946 & -44.5291328 & -1.0000 &  3.15$\pm$0.16 &   7.5$\pm$  0.9 &  28.59$\pm$  2.14 & B  \\
  68.b\mb & 342.1711426 & -44.5342445 &    -    &         -        &   5.6$\pm$  0.5 &  30.02$\pm$  1.41 & B  \\
  68.c\mb & 342.1867371 & -44.5427361 &    -    &         -        &   2.6$\pm$  0.3 &   0.00$\pm$  0.00 & B  \\
\hline
  69.a\mb & 342.1822510 & -44.5260963 & -1.0000 & 12.65$\pm$6.93 &  42.1$\pm$ 12.6 & 124.38$\pm$  1.46 & B  \\
  69.b\mb & 342.1798096 & -44.5256767 &    -    &         -        &  60.9$\pm$ 40.5 & 126.03$\pm$  2.01 & B  \\
  69.c\mb & 342.1981506 & -44.5428619 &    -    &         -        &   1.4$\pm$  0.0 &   0.00$\pm$  0.00 & B  \\
\hline
  70.a\mb & 342.1977234 & -44.5194969 & -1.0000 &  6.49$\pm$0.38 & 136.8$\pm$ 71.9 &   9.34$\pm$  2.52 & B  \\
  70.b\mb & 342.2059631 & -44.5292969 &    -    &         -        &   3.1$\pm$  0.3 &   0.00$\pm$  0.00 & B  \\
\hline
  71.a\mb & 342.1837158 & -44.5383453 & -1.0000 &  3.63$\pm$0.42 &  20.1$\pm$  1.7 &  71.17$\pm$  1.11 & B  \\
  71.b\mb & 342.1807556 & -44.5375061 &    -    &         -        &  21.6$\pm$  1.3 &  71.15$\pm$  0.83 & B  \\
  71.c\mb & 342.1694031 & -44.5216217 &    -    &         -        &   2.2$\pm$  0.0 &   0.00$\pm$  0.00 & B  \\
\hline
  72.a\mb & 342.2002869 & -44.5347557 & -1.0000 &  3.44$\pm$0.41 &   2.8$\pm$  0.1 &   0.00$\pm$  0.00 & B  \\
  72.b\mb & 342.1936035 & -44.5236244 &    -    &         -        &   6.5$\pm$  0.7 &  20.89$\pm$  1.28 & B  \\
  72.c\mb & 342.1854858 & -44.5196877 &    -    &         -        &  57.9$\pm$ 82.0 &  17.41$\pm$  1.88 & B  \\
\hline
  73.a\mb & 342.1912231 & -44.5310135 & -1.0000 &  0.95$\pm$0.03 &  17.0$\pm$  0.6 &  11.62$\pm$  0.54 & B  \\
  73.b\mb & 342.1907654 & -44.5300140 &    -    &         -        &  42.5$\pm$  2.0 &  11.74$\pm$  0.44 & B  \\
  73.c\mb & 342.1817017 & -44.5234184 &    -    &         -        &   4.1$\pm$  0.4 &   0.00$\pm$  0.00 & B  \\
\hline
  74.a\mb & 342.1906433 & -44.5252228 & -1.0000 &  3.27$\pm$0.12 &   4.2$\pm$  0.4 &  33.53$\pm$  1.21 & B  \\
  74.b\mb & 342.1797180 & -44.5210762 &    -    &         -        &   5.2$\pm$  1.0 &  24.11$\pm$  2.54 & B  \\
  74.c\mb & 342.1974487 & -44.5373573 &    -    &         -        &   2.7$\pm$  0.1 &   0.00$\pm$  0.00 & B  \\
\hline
  75.a\mb & 342.1966553 & -44.5375061 & -1.0000 &  3.41$\pm$0.05 &   2.9$\pm$  0.1 &   0.00$\pm$  0.00 & B  \\
  75.b\mb & 342.1907349 & -44.5258865 &    -    &         -        &   3.8$\pm$  0.3 &  30.16$\pm$  1.33 & B  \\
  75.c\mb & 342.1780701 & -44.5209885 &    -    &         -        &   4.6$\pm$  0.8 &  14.27$\pm$  2.61 & B  \\
\hline
  76.a\mb & 342.1903687 & -44.5342255 & -1.0000 &  2.95$\pm$0.23 &  32.1$\pm$ 15.3 &  79.14$\pm$  1.13 & B  \\
  76.b\mb & 342.1900940 & -44.5318031 &    -    &         -        &  25.1$\pm$  6.8 &  79.11$\pm$  1.20 & B  \\
  76.c\mc & 342.1729126 & -44.5195541 &    -    &         -        &   2.3$\pm$  0.1 &   0.00$\pm$  0.00 & B  \\
\hline
  77.a\md & 342.1841125 & -44.5244598 & -1.0000 &  3.24$\pm$3.57 &  29.9$\pm$  4.1 &   0.00$\pm$  0.00 & B  \\
  77.b\md & 342.1821289 & -44.5240440 &    -    &         -        &  25.1$\pm$  8.3 &   0.20$\pm$  0.08 & B  \\
\hline
  78.a\md & 342.1871948 & -44.5236969 & -1.0000 & 14.00$\pm$2.19 &   8.1$\pm$  1.1 &  79.67$\pm$  2.43 & B  \\
  78.b\md & 342.1813354 & -44.5221901 &    -    &         -        &  12.0$\pm$  4.5 &  79.86$\pm$  3.57 & B  \\
  78.c\md & 342.1989136 & -44.5401039 &    -    &         -        &   1.8$\pm$  0.1 &   0.00$\pm$  0.00 & B  \\
\hline
  79.a\md & 342.1756592 & -44.5293427 & -1.0000 &  1.73$\pm$1.13 &  30.0$\pm$  3.5 &   0.19$\pm$  0.07 & B  \\
  79.b\md & 342.1758118 & -44.5299263 &    -    &         -        & 199.9$\pm$  0.0 &   0.00$\pm$  0.00 & B  \\
\hline
  80.a\me & 342.1820984 & -44.5202484 & -1.0000 & 14.93$\pm$1.64 &   7.3$\pm$  2.1 &  49.31$\pm$  4.00 & B  \\
  80.b\me & 342.1909180 & -44.5232277 &    -    &         -        &   5.0$\pm$  0.6 &  51.94$\pm$  2.73 & B  \\
  80.c\me & 342.2002258 & -44.5380516 &    -    &         -        &   2.1$\pm$  0.1 &   0.00$\pm$  0.00 & B  \\
\hline
  81.a\me & 342.2159729 & -44.5194283 & -1.0000 &  2.57$\pm$0.41 &   3.7$\pm$  1.2 &   1.68$\pm$  0.27 & B  \\
  81.b\me & 342.2154236 & -44.5191040 &    -    &         -        &   4.2$\pm$  0.5 &   1.68$\pm$  0.23 & B  \\
  81.c\me & 342.2141113 & -44.5168419 &    -    &         -        &   2.4$\pm$  0.7 &   0.00$\pm$  0.00 & B  \\
\hline
  82.a\me & 342.2151794 & -44.5192299 & -1.0000 &  0.92$\pm$0.05 &   3.2$\pm$  0.7 &   0.30$\pm$  0.02 & B  \\
  82.b\me & 342.2147827 & -44.5188599 &    -    &         -        &   1.8$\pm$  0.5 &   0.16$\pm$  0.04 & B  \\
  82.c\me & 342.2144165 & -44.5180359 &    -    &         -        &   2.7$\pm$  0.5 &   0.00$\pm$  0.00 & B  \\
\hline
  83.a\me & 342.2166443 & -44.5164528 & -1.0000 & 16.65$\pm$0.03 &   2.9$\pm$  1.0 &   0.00$\pm$  0.00 & B  \\
  83.b\me & 342.2179565 & -44.5187454 &    -    &         -        &  40.7$\pm$ 23.4 &   0.25$\pm$  0.09 & B  \\
  83.c\me & 342.2178345 & -44.5183487 &    -    &         -        &   3.8$\pm$  0.4 &   0.15$\pm$  0.07 & B  \\
\hline
  84.a\me & 342.2216492 & -44.5202942 & -1.0000 &  1.16$\pm$0.43 &   3.5$\pm$  0.4 &   0.00$\pm$  0.00 & B  \\
  84.b\me & 342.2217407 & -44.5198936 &    -    &         -        &   1.2$\pm$  0.3 &   0.04$\pm$  0.01 & B  \\
  84.c\me & 342.2215881 & -44.5196037 &    -    &         -        &   2.3$\pm$  0.3 &   0.08$\pm$  0.01 & B  \\
\hline
  85.a\mf & 342.2150879 & -44.5193481 & -1.0000 &  1.04$\pm$0.26 &   2.9$\pm$  0.7 &   0.00$\pm$  0.01 & B  \\
  85.b\mf & 342.2145386 & -44.5187798 &    -    &         -        &   3.1$\pm$  0.9 &   0.13$\pm$  0.05 & B  \\
  85.c\mf & 342.2143555 & -44.5183449 &    -    &         -        &   8.6$\pm$  0.9 &   0.03$\pm$  0.04 & B  \\
\hline
  86.a\mf & 342.2092590 & -44.5205498 & -1.0000 & 16.65$\pm$5.00 &  14.9$\pm$  5.8 &   0.00$\pm$  0.00 & B  \\
  86.b\mf & 342.2074585 & -44.5192032 &    -    &         -        &  37.8$\pm$  2.1 &   0.23$\pm$  0.05 & B  \\
\hline
  87.a\mf & 342.2159729 & -44.5038414 & -1.0000 & 16.65$\pm$0.07 &   3.2$\pm$  0.3 &   0.00$\pm$  0.00 & B  \\
  87.b\mf & 342.2154541 & -44.5035019 &    -    &         -        &   2.6$\pm$  0.2 &   0.01$\pm$  0.01 & B  \\
\hline
  88.a\mf & 342.2066345 & -44.5122414 & -1.0000 &  0.84$\pm$0.09 &   0.7$\pm$  0.2 &   1.79$\pm$  0.27 & B  \\
  88.b\mf & 342.2050781 & -44.5112762 &    -    &         -        &   1.5$\pm$  0.2 &   0.00$\pm$  0.00 & B  \\
\hline
  89.a\mf & 342.1987915 & -44.5196228 & -1.0000 &  2.14$\pm$1.32 &  61.9$\pm$ 79.7 &   0.09$\pm$  0.06 & B  \\
  89.b\mf & 342.1992493 & -44.5200615 &    -    &         -        &  26.6$\pm$  5.0 &   0.16$\pm$  0.06 & B  \\
  89.c\mf & 342.1999207 & -44.5206718 &    -    &         -        &  80.3$\pm$ 48.1 &   0.35$\pm$  0.15 & B  \\
  89.d\mf & 342.2004700 & -44.5212288 &    -    &         -        &  36.7$\pm$ 18.9 &   0.12$\pm$  0.20 & B  \\
  89.e\mf & 342.2011719 & -44.5218735 &    -    &         -        & 105.5$\pm$ 69.7 &   0.31$\pm$  0.21 & B  \\
  89.f\mf & 342.2017517 & -44.5225258 &    -    &         -        & 111.3$\pm$ 65.1 &   0.14$\pm$  0.05 & B  \\
\hline
  90.a\mf & 342.2119446 & -44.5218964 & -1.0000 & 16.65$\pm$0.03 &   5.2$\pm$  1.8 &   0.15$\pm$  0.06 & B  \\
  90.b\mf & 342.2116699 & -44.5214539 &    -    &         -        &   5.4$\pm$  0.3 &   0.10$\pm$  0.02 & B  \\
  90.c\mf & 342.2114563 & -44.5212708 &    -    &         -        &   7.4$\pm$  2.8 &   0.00$\pm$  0.00 & B  \\
  90.d\mf & 342.2107239 & -44.5204544 &    -    &         -        &  12.3$\pm$  5.1 &   0.25$\pm$  0.01 & B  \\
\hline
  91.a\mf & 342.1986389 & -44.5227509 & -1.0000 &  1.47$\pm$0.49 &  63.3$\pm$ 43.9 &   0.05$\pm$  0.03 & B  \\
  91.b\mf & 342.1993408 & -44.5234871 &    -    &         -        &  39.6$\pm$ 10.5 &   0.14$\pm$  0.09 & B  \\
  91.c\mf & 342.1977234 & -44.5219269 &    -    &         -        &  31.6$\pm$  6.1 &   0.00$\pm$  0.00 & B  \\
\hline
  92.a\mf & 342.1630554 & -44.5385437 & -1.0000 & 16.65$\pm$5.00 &   8.1$\pm$  1.5 &   0.06$\pm$  0.07 & B  \\
  92.b\mf & 342.1629944 & -44.5385017 &    -    &         -        &   8.1$\pm$  1.5 &   0.20$\pm$  0.08 & B  \\
  92.c\mf & 342.1624756 & -44.5378036 &    -    &         -        &   7.1$\pm$  1.2 &   0.00$\pm$  0.00 & B  \\
 
\hline
\end{longtable}

{\bf Notes:}
\ma   New counterimage identified in JWST and consistent with lens model
\mb   Full new family of counterimages identified in JWST and consistent with lens model
\mc   Not observed in JWST data (MUSE line or too faint). Position predicted by lens model
\md   Outside JWST footprint but previous identified in HST (BUFFALO) data and consistent with lens model
\me   New family of counterimages identified in HST (BUFFALO) data and consistent with lens model
\mf   Giant arc, likely singly lensed, but useful to constrain the lens model (need to be focussed into a small source).

\twocolumn

\end{appendix}

\end{document}